\documentclass[preprint]{aastex}
\usepackage{csvsimple}
\usepackage{amsmath}
\usepackage[titletoc, title]{appendix}
\pdfoutput=1

\shorttitle{Near-Infrared Photometry of KH 15D}
\shortauthors{Arulanantham et al.}

\begin{document}

\title{Seeing Through the Ring: Near-Infrared Photometry of V582 Mon (KH 15D)}

\author{Nicole A. Arulanantham \& William Herbst}
\affil{Astronomy Department, Wesleyan University, Middletown, CT 06459}
\author{Ann Marie Cody}
\affil{NASA Ames Research Center, Moffett Field, CA 94035}
\author{John R. Stauffer \& Luisa M. Rebull}
\affil{Spitzer Science Center, California Institute of Technology, Pasadena, CA 91125}
\author{Eric Agol \& Diana Windemuth}
\affil{Department of Astronomy, University of Washington, Seattle, WA 98195-1580}
\author{Massimo Marengo}
\affil{Department of Physics and Astronomy, Iowa State University, Ames, IA 50011}
\author{Joshua N. Winn}
\affil{Department of Physics and Kavli Institute for Astrophysics and Space Research, Massachusetts Institute of Technology, Cambridge, MA 02139}
\author{Catrina M. Hamilton}
\affil{Physics and Astronomy Department, Dickinson College, 
        Carlisle, PA 17013}
\author{Reinhard Mundt}
\affil{Max Planck Institute for Astronomy, D-69117 Heidelberg,
    Germany}
\author{Christopher M. Johns-Krull}
\affil{Department of Physics and Astronomy, Rice University, Houston, 
       TX 77005}
\and
\author{Robert A. Gutermuth}
\affil{Department of Astronomy, University of Massachusetts, Amherst, MA 01002}

\author{}
\author{}

\begin{abstract}

We examine the light and color evolution of the T Tauri binary KH 15D through photometry obtained at wavelengths between 0.55 and 8.0 $\mu$m. The data were collected with ANDICAM on the 1.3 m SMARTS telescope at Cerro-Tololo Inter-American Observatory and with IRAC on the Spitzer Space Telescope. We show that the system's circumbinary ring, which acts as a screen that covers and uncovers different portions of the binary orbit as the ring precesses, has reached an orientation where the brighter component (star B) fully or nearly fully emerges during each orbital cycle. The fainter component (star A) remains fully occulted by the screen at all phases. The leading and trailing edges of the screen move across the sky at the same rate of $\sim$15 meters per second, consistent with expectation for a ring with a radius and width of $\sim$4 AU and a precession period of  $\sim$6500 years. Light and color variations continue to indicate that the screen is sharp edged and opaque at \emph{VRIJH} wavelengths. However, we find an increasing transparency of the ring edge at 2.2, 3.6, and 4.5 $\mu$m. Reddening seen at the beginning of the eclipse that occurred during the CSI 2264 campaign particularly suggests selective extinction by a population of large dust grains. Meanwhile, the gradual bluing observed while star B is setting is indicative of forward scattering effects at the edge of the ring. The SED of the system at its bright phase shows no evidence of infrared excess emission that can be attributed to radiation from the ring or other dust component out to 8 microns. 

\end{abstract}

\keywords{stars: individual (KH15D) --- stars: pre-main sequence --- protoplanetary disks --- Galaxy: open clusters and associations: individual (NGC 2264)}

\section{Introduction}

\begin{figure}
\epsscale{1.0}
\plotone{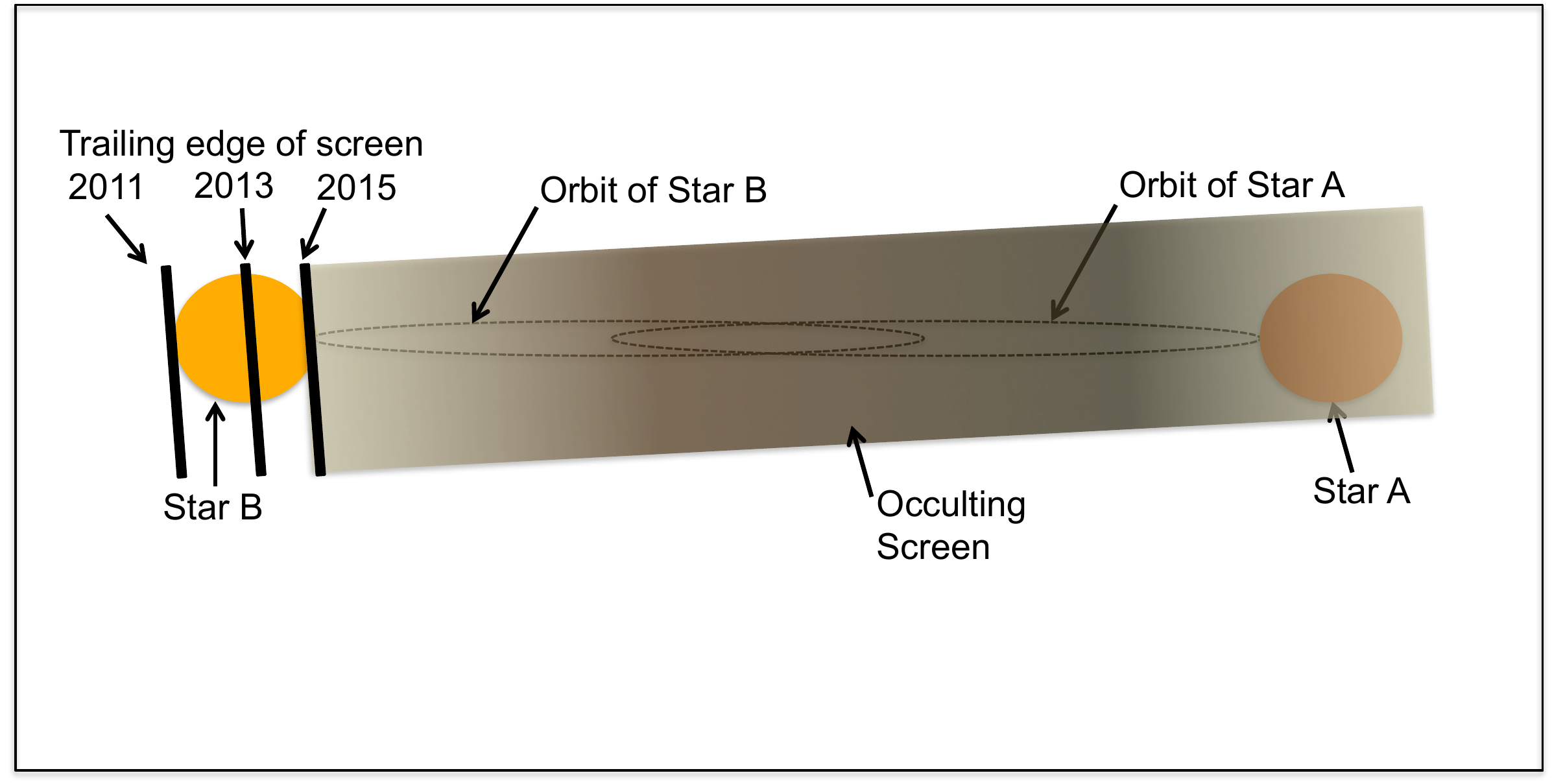}
\caption{The stellar components of KH 15D have spectral types K6/K7 (star A) and K1 (star B). The stars are surrounded by a narrow, rigid circumbinary ring that is inclined with respect to the orbital paths of the stars and precesses over a timescale of $\sim$6500 years. The projection of the ring on the sky acts as an occulting screen, which moves to the right in the diagram above as the ring precesses. The screen is currently positioned such that the entire orbit of star A is covered at all times. Star B becomes visible near apastron, when the system is brightest.}
\label{cartoon}
\end{figure}

V582 Mon, more commonly known as KH 15D, is a binary system located in the $\sim$3 Myr old open cluster NGC 2264 at a distance of 760 pc \citep{S97}. The object was included in a survey of young variable objects and noted because of the unusual depth and duration of its strictly periodic brightness variations \citep{KH98}, which were later found to be evolving in a manner never before seen in any astronomical object \citep{H03}. Models of the system \citep{CM04, W04} explained its behavior by placing two stars at the center of a warped, rigidly precessing ring of material that roughly stretches from $\sim$1-5 AU. See Figure \ref{cartoon} for a sketch of the geometry. The projection of the circumbinary ring on the sky creates an opaque, sharp-edged screen, and the system brightens and fades as the stars appear to ``rise" and ``set" at its edge. Spectral analysis confirmed that KH 15D is a binary with an orbital period of 48.37 days \citep{J04, H05}. The orbital period is equal to the period over which the observed light from the system varies on the short term, confirming that the motion of the stars causes the eclipses. Both components are classified as weak-lined T Tauri stars (WTTS), based on the system's H-$\alpha$ equivalent width of 2 $\AA$ at maximum light \citep{H03}. However, the strength of the H-$\alpha$ line increases to 40 $\AA$ at mid-eclipse and 60 $\AA$ at egress. The system is also still actively accreting, as evidenced by a bipolar jet \citep{T04, D04} and forbidden emission lines \citep{M10}. Analyzing the structure and composition of the circumbinary ring may shed light generally on the earliest stages of terrestrial planet formation in low mass stars, particularly low mass binaries. 

The inner radius of the dust ring is presumably truncated sharply by the time variable gravitational potential of the binary, although gas gets through to cause accretion and jet activity. Why and how the ring is truncated at its outer edge is less clear. Color excesses in \emph{I-J} and \emph{I-H} have been detected, and the amount of excess is consistent with the flux expected from a $\sim$1 Myr-old, 10 $M_{Jupiter}$ planet \citep{WH14}. If spectroscopically confirmed, this could be the signature of a shepherding planet that constrains the ring's outer radius. Regardless of what truncates it, the ring precesses rigidly \citep{CM04}, causing the system's eclipse properties to vary as the screen covers different portions of the binary orbit over time \citep{W03,JW04}. The less massive star (star A) was the only visible component from 1995-2009 and emerged from behind the screen at each apastron passage during that era. Both stars were covered from 2009 to 2011, but the trailing edge of the screen began to uncover the orbit of the brighter companion (star B) at apastron passages during late 2011, while star A remained completely hidden \citep{C12}. In this paper we present two more observing seasons (2013 - 2015) of optical and near-IR photometry covering a period of continued unveiling of star B and, therefore, brightening of the system at all phases, most noticeably at maximum light.  

We also present infrared data from the Spitzer Space Telescope at several epochs from 2004 - 2013 and covering all phases of the orbital cycle, to extend our analysis of the system's properties to longer wavelengths. We compare the Spitzer magnitudes with the optical data at the appropriate epoch to determine if the ring remains optically thick at these wavelengths and to search for excess emission from the ring, disk or putative giant planet. Light curves in the \emph{VRIJH} bands have shown eclipse depths of $\thicksim$4 magnitudes, indicating sharp attenuation of starlight by the ring material even near its edge. The system's \emph{VRIJH} colors show no reddening as star B ``sets". In fact, the colors at these wavelengths become somewhat bluer as the star disappears completely, a feature we attribute to forward scattering by grains of a size similar to the wavelength of light \citep{SA08}. Although little is known about the nature of the material itself, polarization measurements indicate nearly achromatic scattering and suggest a population of grains with a size on the order of $\thicksim$10 $\mu$m \citep{A04}.

We first discuss the latest developments in the evolving light curve, as well as our interpretation of the Spitzer photometry. We also emphasize the Spitzer data reduction process, in which careful correction for contamination by diffracted light from NGC 2264 IRS1 and emission from the inner jet of the system is required. This paper benefitted enormously from the fact that KH 15D was included in the fields monitored during the ``Coordinated Synoptic Investigation of NGC 2264" (CSI 2264) program \citep{C14}, providing us with very high cadence data over a 28-day span in Dec. 2011, which coincided with a partial egress, bright phase, and partial ingress of star B. 

\section{Data and Reductions}

\subsection{ANDICAM Observations}

\begin{figure}
\epsscale{1.0}
\plotone{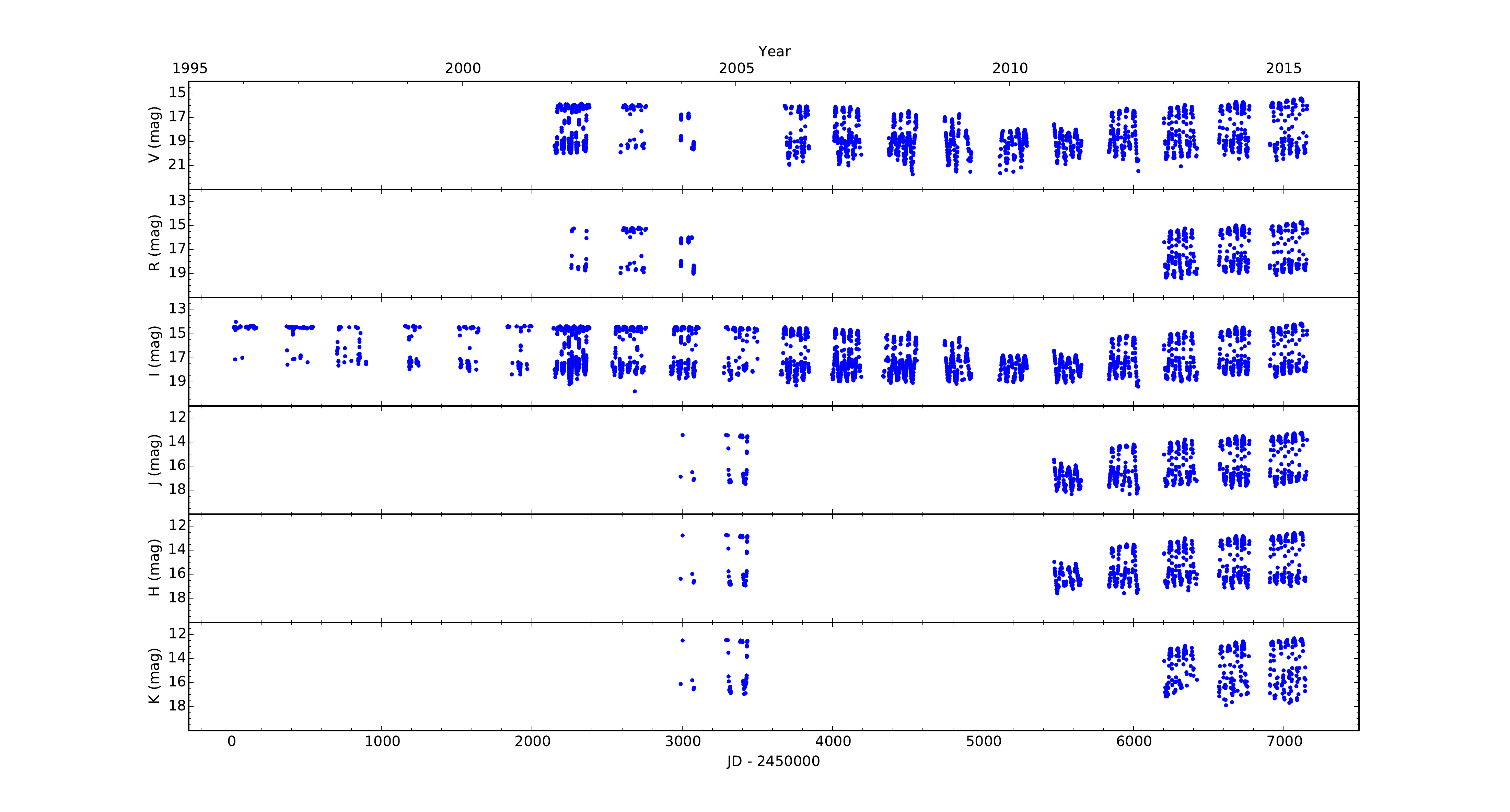}
\caption{All CCD photometry of KH 15D obtained since 1995 (see Table \ref{VRIJHKData}) showing how the system has evolved with time. Prior to 2006, Star A was fully visible during each apastron passage. From 2006 until 2010, the leading edge of the precessing disk was cutting through the photosphere of star A at each apastron passage, causing the system to become progressively fainter at maximum light. The full orbit of both stars was occulted in 2010 and 2011, leaving only scattered light from the ring edges to observe. In 2012, star B emerged from the trailing edge of the ring and the system has progressively brightened at each apastron passage since then. We are currently at or close to the point where star B is fully rising during each cycle.}
\label{longterm}
\end{figure}

We have continued to obtain ground-based optical and near-infrared photometry over the last two years using A Novel Dual Imaging CAMera (ANDICAM) on the 1.3 m telescope at Cerro-Tololo Inter-American Observatory (CTIO) in Chile. The instrument is operated by the SMARTS consortium. Data were collected almost nightly from October 2013 through April 2014. Observations were resumed in September 2014 and continued until April 2015. Each night, four 150 s exposures were obtained in each of the three optical bands $\left(VRI\right)$ along with 10-15 dithered exposures (30 s each) in the near-infrared bands $\left(JHK\right)$. All images have a 10.2$'$ x 10.2$'$ field of view. The data acquisition and reduction processes are discussed briefly in Appendix A, and a more complete description is given by \citet{WH14}. The $VRIJHK$ magnitudes from the last two observing seasons have been added to the entire set of CCD data obtained since 1995, which is presented here as Table \ref{VRIJHKData}. The 14,280 data points are plotted versus time in Figure \ref{longterm}. 

\begin{deluxetable}{lllllllllllllll}
\tablewidth{0 pt}
\tabletypesize{\scriptsize}
\tablecaption{$VRIJHK$ Photometry of KH 15D \label{VRIJHKData}}
\tablehead{
\colhead{JD} & \colhead{V} & \colhead{$\sigma_V$} & \colhead{R} & \colhead{$\sigma_R$} & \colhead{I} & \colhead{$\sigma_I$} & \colhead{J} & \colhead{$\sigma_J$} & \colhead{H} & \colhead{$\sigma_H$} & \colhead{K} & \colhead{$\sigma_K$} & \colhead{Obs\tablenotemark{a}} & \colhead{Ref\tablenotemark{b}}
}
\startdata
 2457140.48 & - & - & - & - & 17.81 &  0.066 & 16.97 & 0.080 & - & - & - &  - &      CTIO &     A15 \\
2457140.50 & - &  - & - &  - & - & - & - & -  &  - & - & 16.28 &       0.349 &      CTIO &     A15 \\
2457141.49 & 19.76 &       0.294 & 18.68 &       0.138 & 18.06 &       0.085 & 16.80 &       0.080 & 16.40 &       0.095 & 16.71 &       0.547 &      CTIO &     A15 \\
2457142.48 & 19.66 &       0.245 & 18.72 &       0.151 & 18.07 &       0.107 & 16.92 &       0.082 & 16.56 &       0.127 & 14.74 &       0.140 &      CTIO &     A15 \\
2457143.51 & 19.91 &       0.390 & 18.45 &       0.132 & 18.12 &       0.117 & 17.10 &       0.103 & 16.27 &       0.091 & 15.75 &       0.202 &      CTIO &     A15 \\
\enddata
\tablecomments{The full table is available electronically. Here we show a sample. In the electronic table, missing magnitudes are coded as 99.99 and missing errors are coded as 9.999.}
\tablenotetext{a}{Observatory and/or Instrument where data were obtained:\\
IRSF = 1.4 m Infrared Survey Facility telescope \\
ESO = European Southern Observatory, La Silla, Chile \\
CTIO = Cerro Tololo Inter-American Observatory SMARTS 1.3 m telescope ANDICAM \\
KAIT = Katzman Automated Telescope, Mt. Hamilton, CA \\
KONK = Konkoly Observatory, Hungary \\
KPNO = Kitt Peak National Observatory telescopes \\
TEN = Tenagra Observatory, Nogales, AZ \\
TENE = Teide Observatory, Tenerife, Spain \\
USNO = US Naval Observatory's 1.3m and 1m telescopes \\
UZBK = Mount Maidanak Observatory, Uzbekistan \\
VVO = Van Vleck Observatory, Wesleyan University, Middletown, CT \\
WISE = Wise Observatory, Israel 
}
\tablenotetext{b}{Code for Paper References to Original Data: \\
H05 = \citet{H05} \\ 
H10 = \citet{H10} \\ 
K05 = \citet{K05} \\
W13 = \citet{WH14} \\
A16 =  Arulanantham et al. (2016) (This paper.)}
\end{deluxetable}

\subsection{Spitzer Data Sets}

Images of KH 15D were collected with the InfraRed Array Camera (IRAC; Fazio et al. 2004) on the Spitzer Space Telescope \citep{Wern04} during six observational runs with five separate PI's spanning three distinct epochs since 2004. Details are given in Table \ref{SpitzSummary}, and all data sets are now publicly available through the Spitzer Heritage Archive. The first four were obtained between 2004 and 2006, when the orbital path of star B was still completely occulted at all phases and star A was mostly or entirely visible during each apastron passage. These observations were conducted during the so-called ``cryogenic" period, when IRAC could be sufficiently cooled to return images from all four of its detectors (3.6, 4.5, 5.8, and 8.0 $\mu$m). Although each of these data sets was reduced at the time it was first obtained by the PIs, the photometry was difficult to interpret because the background flux is comparable to the signal from KH 15D. Results have therefore not been previously published. 

IRAC had exhausted its supply of liquid helium coolant by the time the last two sets of Spitzer images were collected. These data only cover 3.6 and 4.5 $\mu$m, since the shorter wavelength detectors do not require operating temperatures to be as low as their longer wavelength counterparts \citep{F04}. The fifth set of observations was obtained by the CSI 2264 team as part of a large campaign to monitor young variable objects in NGC 2264 \citep{C14}. These data were obtained over 28 consecutive days of observation in December 2011, when star A was completely hidden from view at all phases and star B was partly visible during each apastron passage. In the high cadence CSI 2264 monitoring program, sequential images were separated by less than a tenth of a day on average in both bands, although a few points were collected after $\sim$1 day had elapsed. The data span phases $\sim$0.3-0.9 of a single cycle, a range which includes part of one egress, all of the peak brightness phase (in which star B was still partly occulted) and most of the following ingress, but does not include central eclipse. 

\begin{figure}
\epsscale{1.0}
\plottwo{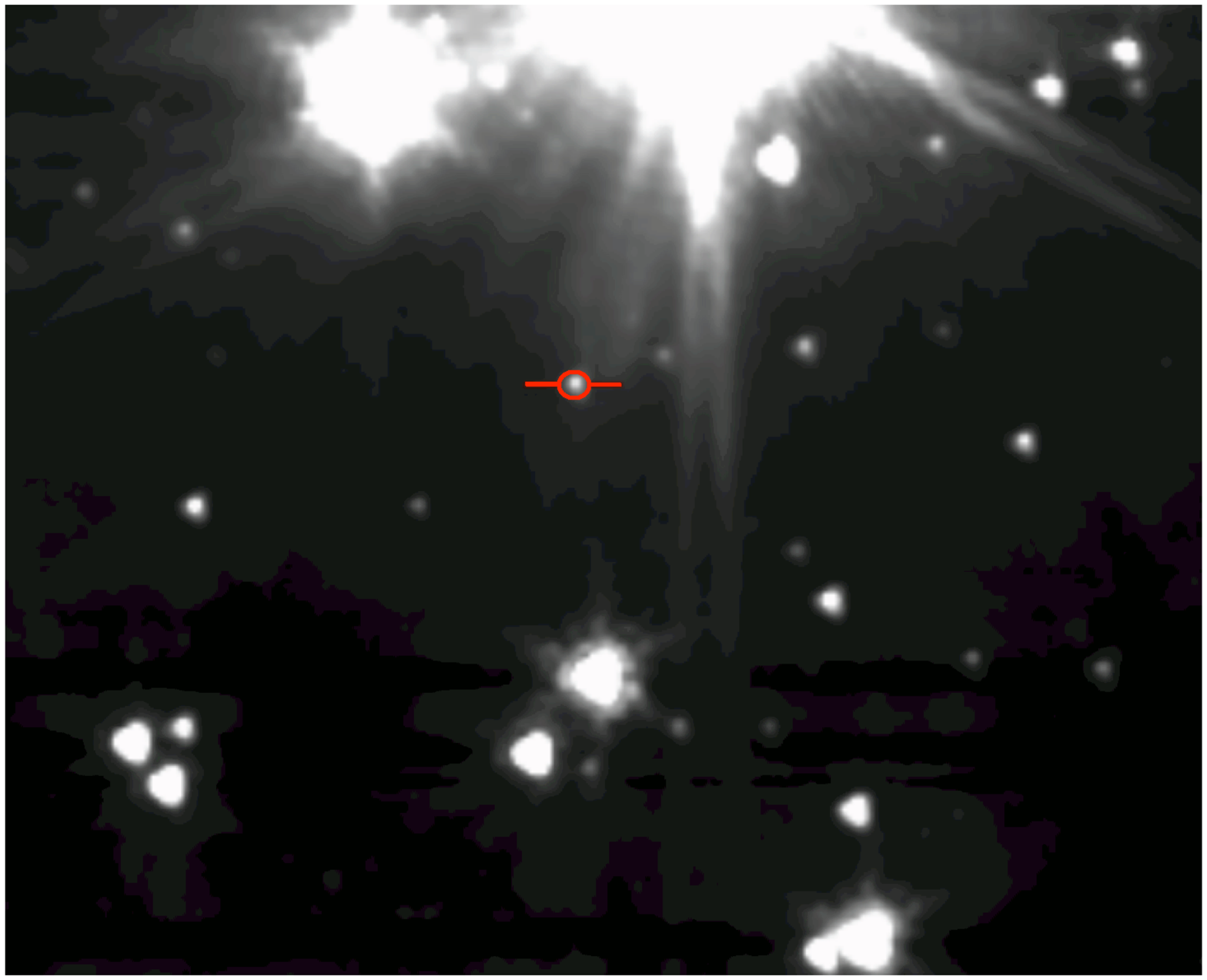}{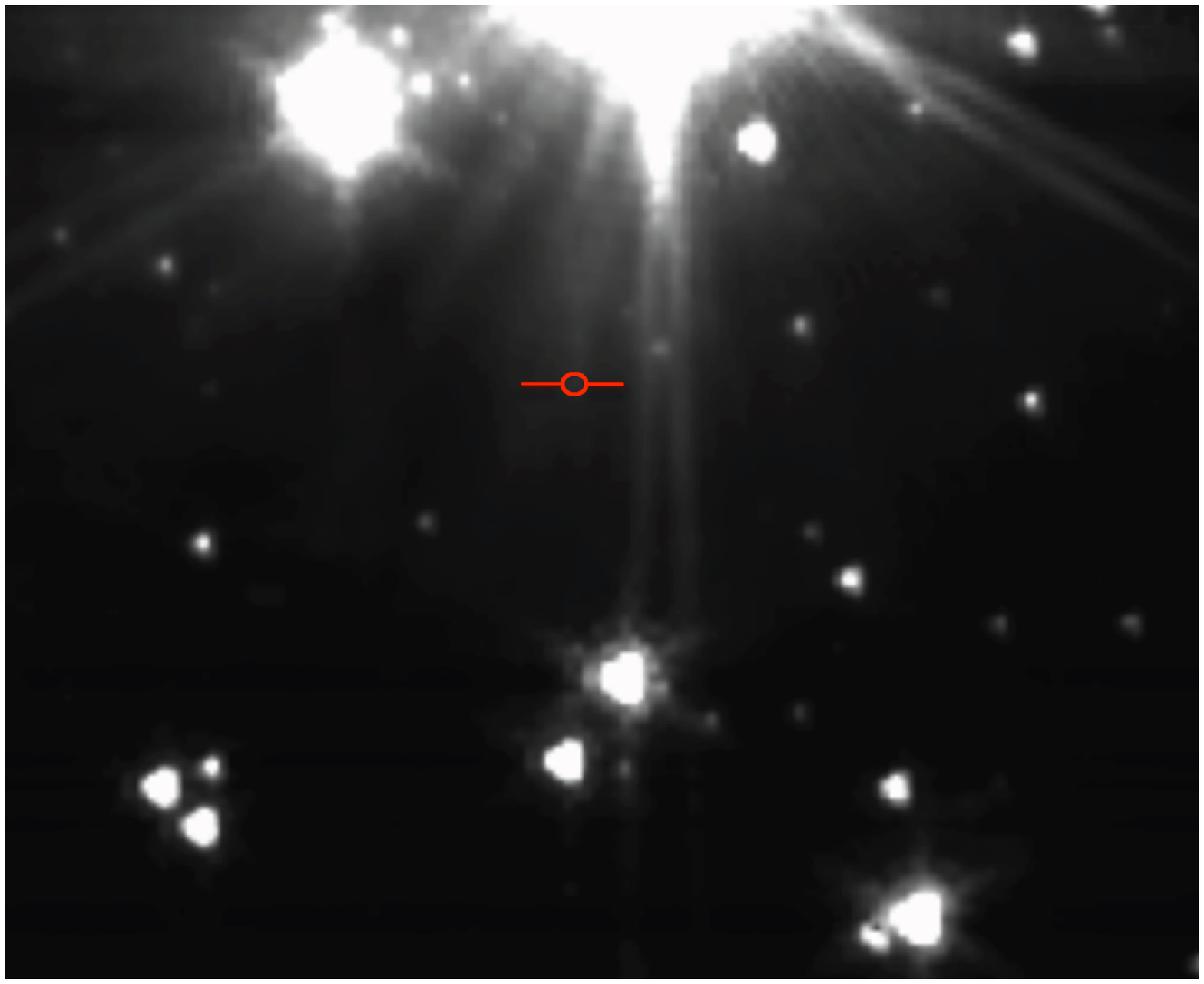}
\caption{Spitzer images in the 3.6 $\mu$m band showing the KH 15D field when the system was near maximum brightness (left) and near minimum brightness (right). North is at the top and East to the left. The size of the region shown is 5.2$\arcmin$ on a side. NGC 2264 IRS1 is about 1$\arcmin$ north of KH 15D (marked by red annuli), and diffraction from the source contaminates the field in the vicinity of KH 15D. The jet from the system, which is not as prominent in these images, extends northward as well.}
\label{BrightFaint}
\end{figure}

A final set of observations was proposed to fill in the gaps in phase space in the 2011 photometry, although the new data would be collected at a much lower cadence. Images were obtained on eight nights between December 2013 and January 2014. A larger fraction of star B was directly visible at apastron by then, although its surface was still partially occulted. These observations covered the central eclipse as well as parts of egress and maximum light. It was found that all of the Spitzer data needed to be reduced with care due to the faintness of the source, especially near periastron passage, compared to contaminating light from two sources -- diffraction spikes from the nearby intense source NGC 2264 IRS1 and the inner jet. The magnitude of the problem is illustrated in Figure \ref{BrightFaint}, which shows a 3.6 $\mu$m Spitzer image of the source near maximum brightness and minimum brightness. Our procedure for dealing with the complex background emission is discussed in Appendix B. The full set of Spitzer photometry at all epochs is given in Table \ref{SpitzerData}.  

\begin{deluxetable}{lllll}
\tabletypesize{\scriptsize}
\tablewidth{0 pt}
\tablecaption{Spitzer Data Sets \label{SpitzSummary}}
\tablehead{
\colhead{UT Date} & \colhead{JD (2450000)} & \colhead{Wavelengths ($\mu$m)} & \colhead{PI} & \colhead{Program ID\tablenotemark{a}}
}
\startdata

2004 Mar 6 & 3070 & 3.6, 4.5, 5.8, 8.0 & Giovanni Fazio & 37 \\
2004 Oct 5-12 & 3283-3290 & 4.5, 5.8, 8.0 & Massimo Marengo & 3441 \\
2004 Oct 08 & 3286 & 3.6, 4.5, 5.8, 8.0 & Giovanni Fazio & 37 \\
2005 Oct 21-29 & 3664-3672 & 4.5, 5.8, 8.0 & Massimo Marengo & 3441 \\
2006 Mar 23-27 & 3817-3821 & 3.6, 4.5, 5.8, 8.0 & Eric Agol & 3469 \\
2008 Nov 1-2 & 4771-4772 & 3.6, 4.5, 5.8, 8.0 & Lucas A. Cieza & 50773 \\
2011 Dec 3-2012 Jan 1 & 5898-5927 & 3.6, 4.5 & John R. Stauffer (CSI 2264) & 61027, 80040 \\
2013 Dec 22-2014 Jan 20 & 6648-6677 & 3.6, 4.5 & William Herbst & 90154, 90098 \\
\enddata
\tablenotetext{a}{Program IDs can be used to retrieve the data from the Spitzer Heritage Archive.}
\end{deluxetable}

\begin{deluxetable}{lllllllll}
\tabletypesize{\scriptsize}
\tablewidth{0 pt}
\tablecaption{Spitzer Photometry of KH 15D \label{SpitzerData}}
\tablehead{
\colhead{JD - 2400000.0} & \colhead{[3.6]} & \colhead{error} & \colhead{[4.5]} & \colhead{error} & \colhead{[5.8]} & \colhead{error} & \colhead{[8.0]} & \colhead{error}
}
\startdata

53283.55 & 14.124 & 0.157 & 14.292 & 0.169 & 13.949 & 0.181 & 13.352 & 0.184 \\
53285.72 & 12.442 & 0.072 & 12.844 & 0.087 & 12.732 & 0.103 & 12.617 & 0.131 \\
53287.24 & 12.335 & 0.068 & - & - & - & - & - & - \\
53287.28 & 12.314 & 0.068 & - & - & - & - & - & - \\
53287.33 & 12.313 & 0.068 & 12.799 & 0.085 & 12.488 & 0.092 & 12.661 & 0.134 \\
53289.17 & 12.339 & 0.069 & 12.786 & 0.084 & 12.684 & 0.101 & 12.650 & 0.133 \\ 
53291.19 & 12.343 & 0.069 & 12.835 & 0.086 & 12.938 & 0.114 & 12.761 & 0.140 \\
53665.12 & 17.581 & 0.772 & - & - & - & - & 13.772 & 0.224 \\
53667.10 & 16.873 & 0.557 & - & - & - & - & - & - \\
53669.52 & 15.547 & 0.302 & 16.307 & 0.429 & 14.457 & 0.229 & 14.194 & 0.272 \\
53670.69 & 14.919 & 0.226 & 15.469 & 0.292 & 14.342 & 0.217 & 14.499 & 0.312 \\
53672.51 & 12.968 & 0.092 & 13.346 & 0.109 & 13.157 & 0.126 & 13.059 & 0.161 \\
53817.50 & 14.557 & 0.191 & 14.495 & 0.186 & 14.091 & 0.194 & 13.307 & 0.180 \\
53819.40 & 12.951 & 0.091 & 13.502 & 0.118 & 13.015 & 0.118 & 12.823 & 0.144 \\
53820.39 & 13.191 & 0.102 & 13.380 & 0.111 & 13.357 & 0.138 & 12.794 & 0.142 \\
53821.84 & 12.996 & 0.093 & 13.207 & 0.103 & 13.138 & 0.125 & 12.738 & 0.139 \\

\enddata
\tablecomments{The full table is available electronically. Here we show a sample. In the electronic table the code 99.99 is used for a missing magnitude measurement and 9.999 for a missing error determination.}
\end{deluxetable}

\section{Results}

\subsection{Long-Term Light Curve Evolution: As the Disk Precesses, Star B Emerges}

\begin{figure}
\epsscale{1.0}
\plotone{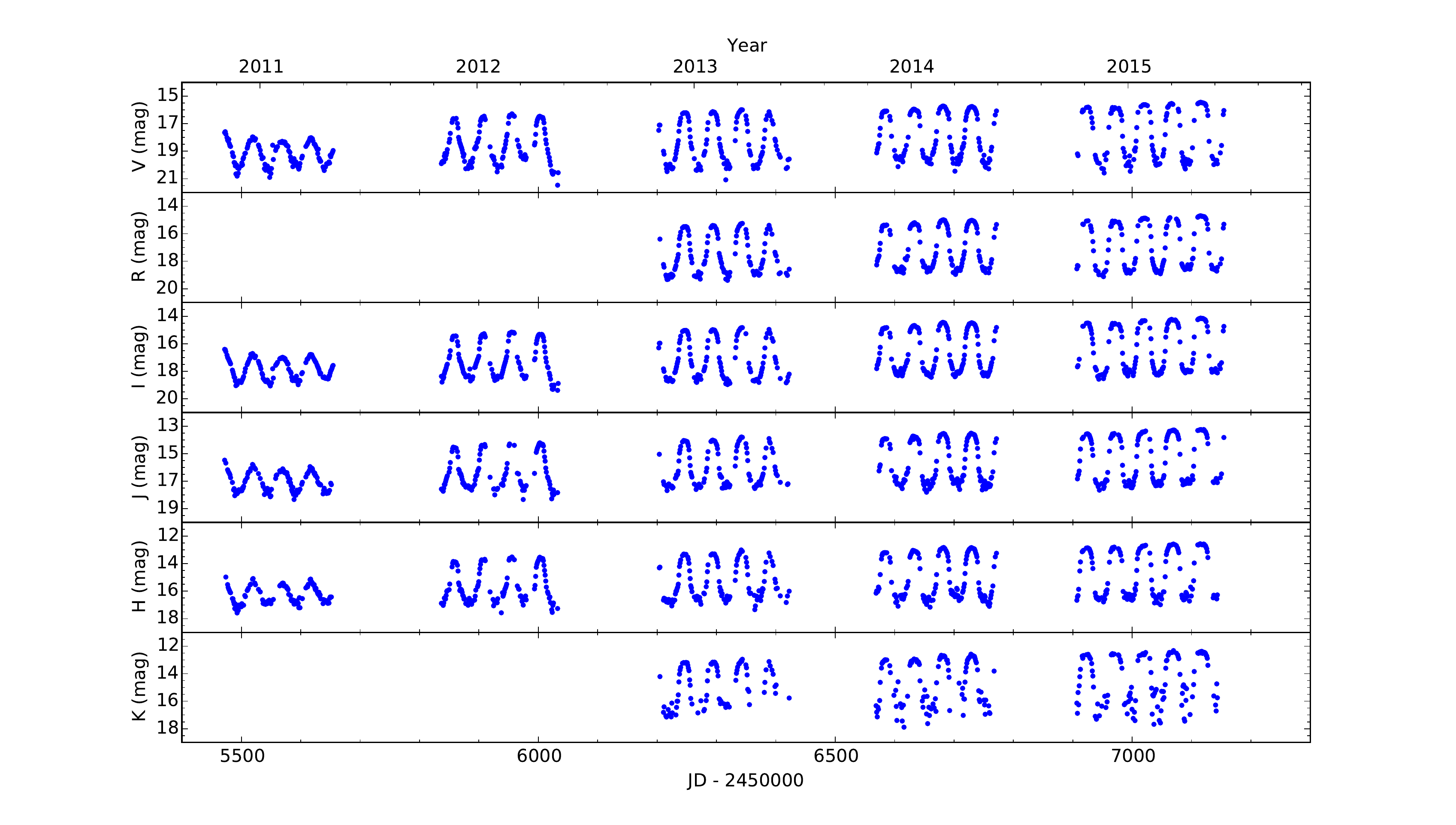}
\caption{The light curves show that the eclipses do not have the same shape from cycle to cycle, nor do they reach precisely the same brightness. This could be caused by irregularities in the circumbinary ring. Cool spots on the surface of the star could also be responsible.}
\label{expandedlightcurve}
\end{figure}

Figure \ref{longterm} places the $VRIJHK$ observations from the last two years in the context of the long-term evolution of the system. We are in a period where the system continues to get brighter at maximum light as more and more of the photosphere of star B (the brighter star in the system) is revealed during each apastron passage. \cite{WH14} predicted that KH 15D would continue to brighten over time and that it would spend increasingly longer time intervals in its bright state as the trailing edge of the occulting screen uncovered more of the orbital path of star B. In Figure \ref{expandedlightcurve}, we show the last five years of data covering the period of full or almost full emergence of Star B. It is clear that the peak magnitude of the system has, in fact, brightened steadily during these observing seasons, confirming our basic understanding of the system. 

Does star B, in fact, fully emerge now at its brightest or is it still at least partially occulted all of the time? The value for the system's expected I-band magnitude when star B is fully emerged can be estimated from the magnitude determined from archival plates obtained between 1965 and 1990 \citep{JW05}, when both stars were directly visible at maximum light ($I = 13.57 \pm 0.06$ mag; Winn et al. 2006), and CCD measurements obtained when star A alone was unocculted $\left(I = 14.47 \pm 0.04 \right)$. This method predicts a peak brightness for the system when star B is fully emerged of of $I=14.19 \pm 0.06$ \citep{WH14}. If this is correct, star B is fully emerged, and we would expect no further increase in brightness of the system near maximum light until both stars become visible at some orbital phase. However, an additional constraint on the unocculted brightness of star B is the photometry obtained at VVO in 1996, when star B was emerging from behind the leading edge of the screen at periastron. The magnitude of the system was measured at $I = 14.01 \pm 0.01$ at this time \citep{KH98}, which is somewhat brighter than the expected value based on the archival data. If the single periastron measurement is a better guide, then we may see some additional brightening of the star next season. At this point we can say that star B is either fully emerging or nearly fully emerging during each orbital cycle.

In either case it is important to note the overall symmetry of the long-term KH 15D light curve (see Figure \ref{longterm}). The radii of stars A and B must be nearly the same, based on their similar brightnesses and temperatures. The fact that the time to cover star A is turning out to be about the same as the time to uncover star B ($\sim$4 years) therefore indicates that the sky projections of the leading and trailing edges of the obscuring ring move at the same rate. Taking the radius of star A to be 1.3 R$_\odot$ \citep{H01} and the time to traverse it as 4 years indicates a velocity across the sky for the projected edge of about 15 m s$^{-1}$. Such a slow, stable progression of both leading and trailing edges is indicative of rigid precession of a ring, as has been assumed now for more than a decade. We return to the ring precession time scale in Section 4.

\subsection{Cycle-to-Cycle Light Curve Variations}

Although Figures \ref{longterm} and \ref{expandedlightcurve} show that KH 15D has steadily been getting brighter each year, the behavior of the light curve is not perfectly uniform among individual cycles. The increase in brightness does not appear to progress at exactly the same rate from cycle-to-cycle and the shapes of the maxima are not always identical. For example, during the 2013-14 observing season, the system got brighter during each of the first three cycles but dimmed by 0.04 mag during the last cycle. The peak magnitude was fainter still for the first two cycles of the 2014-15 season before brightening again by 0.05 mag during each of the last two cycles. The non-uniformity of the cycle-to-cycle light curves could be explained by the presence of hills and valleys located along the occulting edge of the circumbinary ring. Alternatively, cool spots on the surface of star B could sometimes result in less observed flux than expected at peak brightness. If cool spots are the reason for the deviations, we can expect to see redder colors during the cycles when the system became dimmer. Clumps within the ring, on the other hand, would not cause color changes during maximum light, unless they were less opaque at the ANDICAM wavelengths than the bulk of the ring material. The amount of reddening at maximum light could, therefore, make it possible to distinguish between spots and clumps as the source of the variations. 

\cite{H94} found that changes in \emph{V} and \emph{I} magnitudes caused by cool spots on T Tauri stars followed the relation d\emph{V}/d\emph{I}$= 0.7$. For a \emph{V} band variation of 0.03 mag, we can expect a 0.02 mag change in \emph{I} and a color change of $V-I = +0.01$. Between the third and fourth cycles of the 2013-2014 observing season, when the system became 0.03 mag fainter in \emph{V} at maximum light, the \emph{V-I} color at peak brightness reddened by 0.03 mag. The difference in color between the first two cycles in 2014-2015 was $\Delta \left(V-I\right) = 0.01$ mag when the peak \emph{V} mag again became 0.03 mag fainter. Although our observations are consistent with what we would expect for a spotted stellar surface, the \emph{V} band fluctuations aren't large enough to result in significant color changes. The color changes are around the same size as the photometric errors in the data, making it difficult to confirm that the reddening is real. The reddening of the \emph{V-I} and \emph{R-I} colors between the 2013-2014 and 2014-2015 data can be attributed to the continued rise of star B. As more of its surface is revealed during the bright phase, more of its light is directly visible and dominates the contribution from the bluer, forward scattered light. 

In addition to the cycle-to-cycle variations in the system's brightness at maximum light, the shape of the light curve at mid-eclipse also varies between cycles (see Figure \ref{expandedlightcurve}). A few show a bump in the light curve at mid-eclipse, but the shape of the bump is not consistent with the mid-eclipse re-brightenings observed when star A was directly visible at apastron and star B approached the leading edge of the screen at periastron. The slight increases in brightness are not significant when compared to the amount of scatter present in the photometry at minimum light, indicating that scattered light from star A as it approaches the trailing edge of the screen near periastron is not yet contributing significantly to the system's brightness. A likely cause of variations near mid-eclipse is stellar activity associated with the close periapse passage of the two stars, which have a fairly eccentric (e = 0.6) orbit. \citet{H12} have discussed the spectral evidence for increased activity at perihelion passage when the stars are separated by only $\sim$15 stellar radii and their magnetospheres may well interact. 

\subsection{Color Evolution}

\begin{figure}
\epsscale{1.0}
\plotone{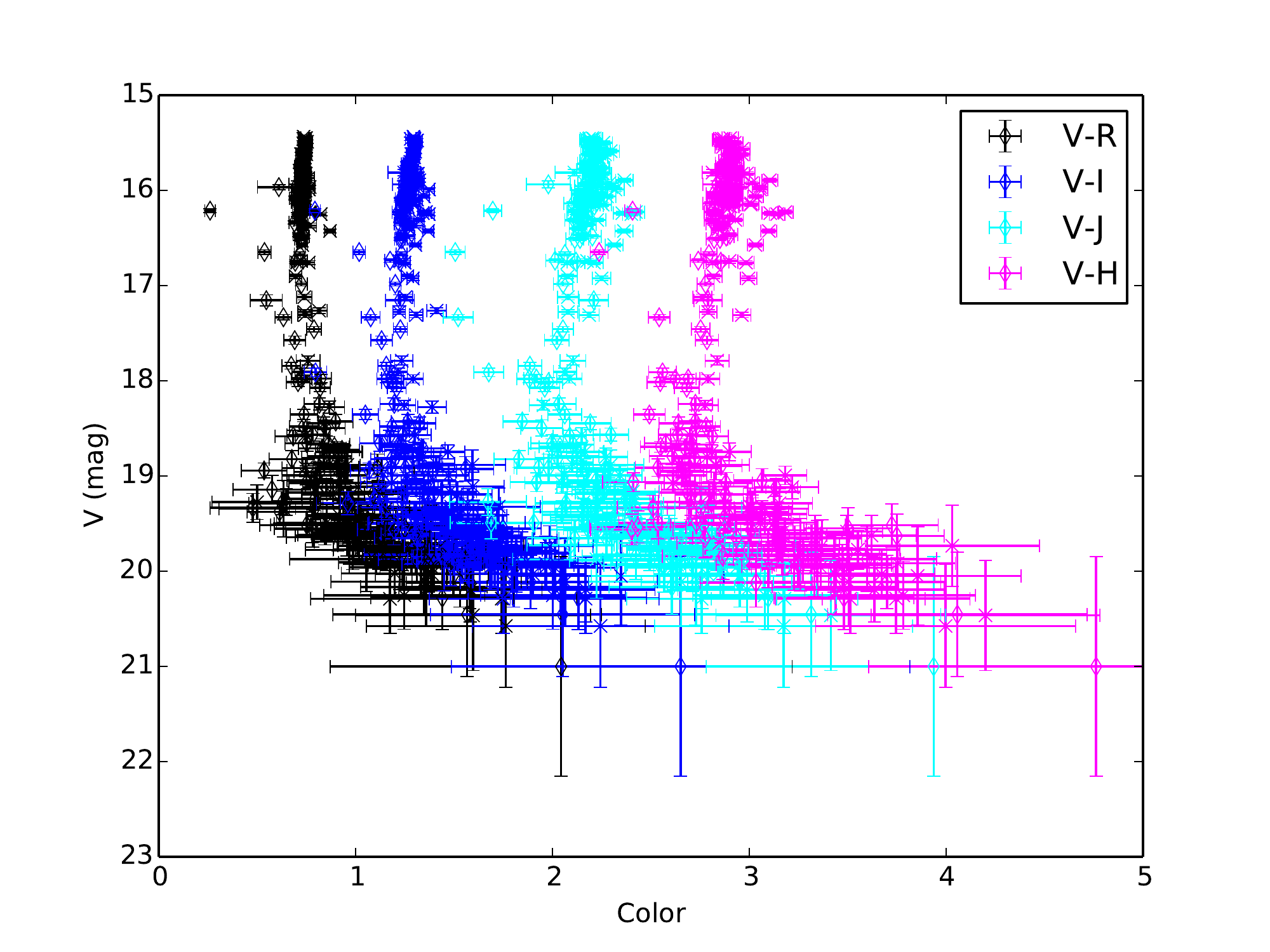}
\caption{Photometry from 2013-2014 is represented by circles and the 2014-2015 data by x's. The system's colors remain fairly constant in \emph{V-R} and \emph{V-I} before reddening dramatically at magnitudes fainter than $V \sim 19$. The transition corresponds to star B's disappearance behind the screen. However, the \emph{V-J} and \emph{V-H} colors become slightly bluer as the system fades from maximum brightness to $V \sim 19$. \cite{SA08} attributed the bluing to forward scattering effects at the edge of the ring; this was confirmed by the forward-scattering model presented in Section 4.2 of this paper.}
\label{ANDICAM_cmd}
\end{figure}

\begin{figure}
\epsscale{0.75}
\plotone{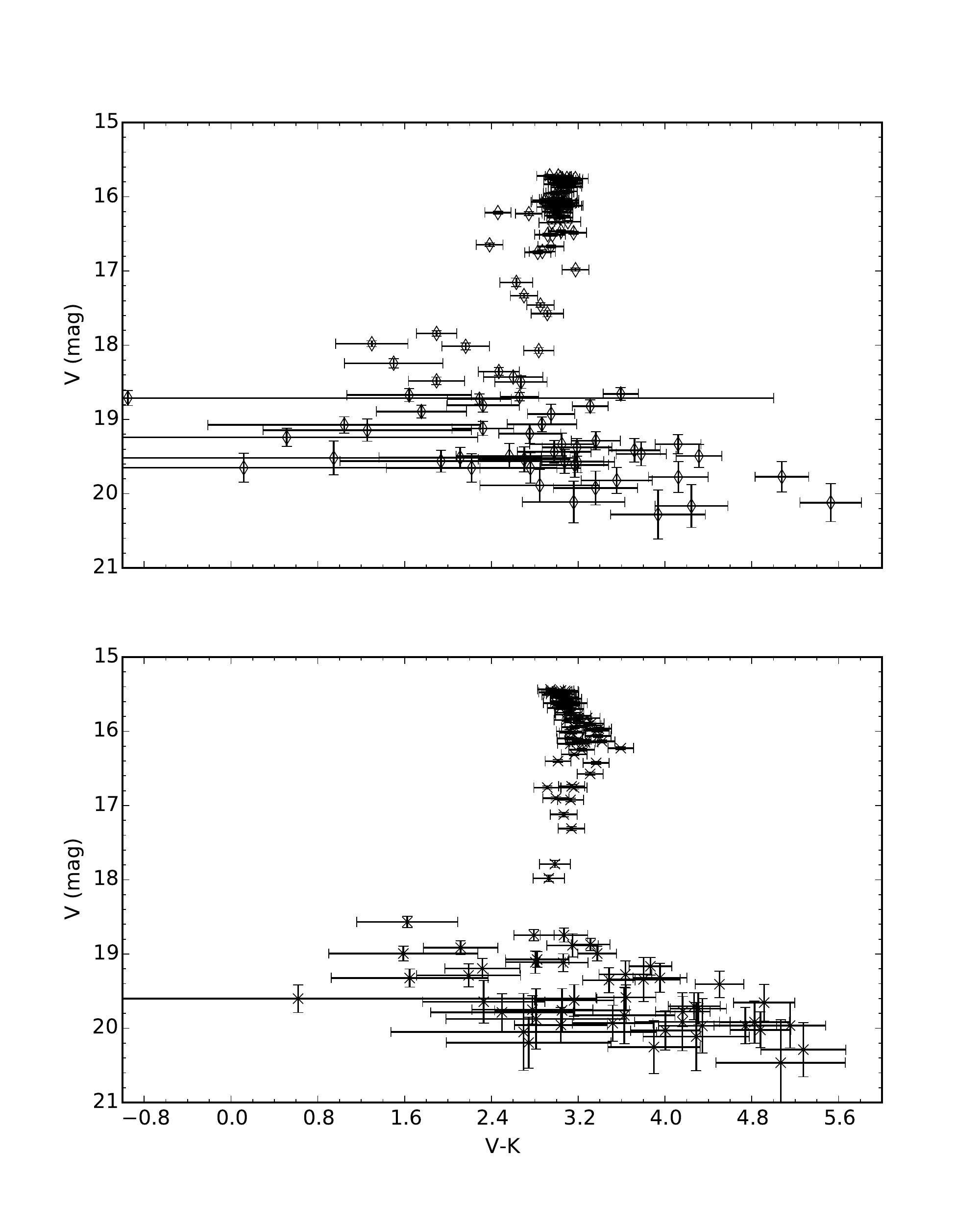}
\caption{Although the \emph{V-K} colors followed the same trend as the shorter wavelengths and stayed constant as the system became fainter in the 2013-2014 data (top), they showed a slight reddening in the 2014-2015 photometry (bottom). The reddening implies that the material currently occulting star B is slightly more transparent than the material that was obscuring it just a year ago.}
\label{VK_cmd_full}
\end{figure}

The color behavior of the KH 15D system as a function of brightness is shown in Figures \ref{ANDICAM_cmd} and  \ref{VK_cmd_full}. Focusing first on the shorter wavelengths, out to H, we see a fairly simple and consistent behavior. At its brightest levels, the light from the system is dominated by photospheric emission from the unocculted portion of star B, and the colors approach values expected for a K1 pre-main sequence star that suffers only a small amount of foreground reddening, $E(B-V) \sim 0.1$ mag. As the system fades due to the setting of star B, scattered light becomes more and more important. Just after ``star set," the light is dominated by forward scattered light off of the ring. The system has a $V$ magnitude of about 18 at this point. It is apparent from Fig. \ref{ANDICAM_cmd} that the color of the scattered light becomes progressively bluer with respect to the photosphere as one moves to longer wavelength baselines. This causes an increasingly large ``tilt" of the color-magnitude curves in the bright phase ($V$ $<$ 18). We attribute the blueness of the scattered light to Mie scattering off of grains with sizes comparable to the wavelength of the light they are scattering; this is supported by a forward scattering model, presented in Section 4.2 of this paper. Since there is actually very little or no bluing of the light at optical wavelengths, it is apparent that the size of the grains must be much larger than is characteristic of the interstellar medium. The scattering material is much ``greyer" than the ISM and is roughly characterized by k$_{scattering} \propto \lambda^{-0.25}$. This nearly achromatic scattering agrees with the results of \citet{A04}, based on spectropolarimetry of the system.  
 
Figure \ref{expandedlightcurve} shows that there is significantly more scatter in the \emph{K} band photometry at minimum light than in the other five ANDICAM bands. \cite{WH14} identified a detection limit of $K \sim 16$ mag due to residual flat-fielding errors and used a smaller aperture for the \emph{K} band photometry than for the \emph{J} and \emph{H} data in order to correct for the lower S/N. Bands \emph{V} through \emph{H} show that KH 15D has now become slightly brighter at all phases since the 2010-2013 photometry was collected, making it brighter than $K \sim 16$ mag during minimum light in the more recent data. If the amount of scatter is a true feature of the \emph{K} band light curve, the ring material could be somewhat more transparent here than at shorter wavelengths.

Figure \ref{VK_cmd_full} shows that the 2014-2015 \emph{V-K} color becomes slightly redder as the system gets fainter, unlike at shorter wavelengths. The trend reverses around $V \sim 16$, as the system returns to the bluer color it showed at peak brightness. This is in contrast to the \emph{V-J} and \emph{V-H} colors (see Figure \ref{ANDICAM_cmd}), which steadily become bluer as star B sets behind the screen and only redden at minimum light. The reddening in the \emph{K} band implies that the opacity of the ring material is lower at $\lambda \sim 2.2 \, \mu$m than at shorter wavelengths. It is also notable that, as we discuss below, there is clear evidence of a similar reddening occurring at the Spitzer wavelengths. As a caution, we do note that this is the first observing season since \emph{K} band observations of KH 15D began in 2011 that such reddening has been detected. We do not have sufficient data to compare the current \emph{V-K} color to the color of the system when star A was the directly visible component and reached a similar peak brightness, but the reddening should be monitored in future observing seasons as the geometry of the system continues to evolve with respect to our line of sight. It could be that our line of sight through the occulting material is leading to a somewhat lower optical depth as time progresses. 

Finally, we note that there are some rather aberrant points visible on the color-magnitude diagrams. We have carefully examined the images on nights when these were detected and could find no reason to exclude the data. Therefore, we have left them on Figures \ref{ANDICAM_cmd} and \ref{VK_cmd_full}. They are particularly noticeable at $J$, $H$ and $K$ wavelengths. It is possible that these are merely accidental photometric errors of some sort, but they could also be actual variations of the system color related to the non-uniformity of the absorbing screen. Not wishing to remove data simply because it does not fit a pattern, we have left those data in the tables and on the figures.  

\subsection{Spitzer Photometry: Seeing through the Ring}

\begin{figure}
\epsscale{0.75}
\plotone{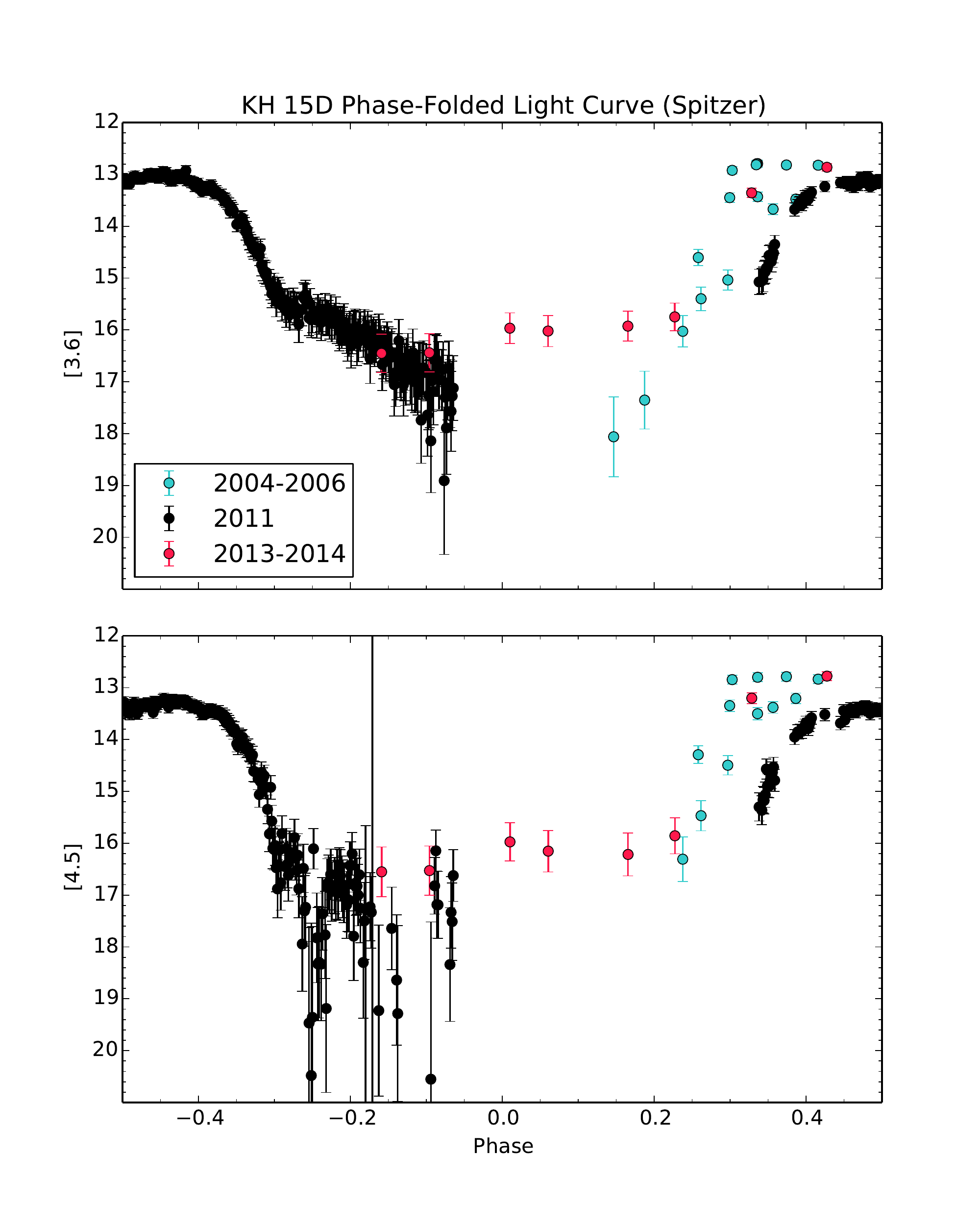}
\caption{The shapes of the phase-folded light curves show that the Spitzer observations were carried out during three distinct epochs. Star A was the only directly visible component when the 2004-2006 data was obtained (light blue). A small fraction of star B was directly visible at apastron during the 2011 observations (black), and more of its surface had emerged by 2013-2014 (red).}
\label{spitzerphasefold}
\end{figure}

The Spitzer photometry was obtained at three distinct epochs, which is evident from the shapes of the 3.6 and 4.5 $\mu$m phased light curves (Figure \ref{spitzerphasefold}). Star A was the only stellar component that was directly visible at the time the 2004-2006 observations were conducted, and its entire surface was unocculted at maximum brightness near apastron. The resulting light curves indicate a shorter eclipse duration and a sharper egress than the other two data sets. The system also rose to a flat maximum, corresponding to those phases when star A was fully visible. Star B began to emerge by 2011, when the system was observed by the CSI 2264 team, although only a small portion of the star was directly visible. The eclipse duration was longer than it was between 2004 and 2006, and the magnitude at maximum brightness was fainter. The fraction of star B that was uneclipsed had increased substantially by the time the 2013-2014 data were collected, corresponding to an increase in peak brightness. Egress was noticeably sharper in the 2011 data but not as abrupt as it was in the 2004-2006 photometry. It was also evident that the full eclipse duration had shortened, although it was still longer than it had been from 2004-2006.

The eclipse depth is more difficult to determine at 3.6 and 4.5 $\mu$m than at shorter wavelengths because of the large amount of scatter in the photometry obtained at minimum light. However, the system had an approximately 3.5 mag decline in brightness over the orbital period during 2011 and 2013-2014. This is consistent with the eclipse depth observed in the \emph{VRIJHK} bands during those years. The eclipse depth may have been deeper at 3.6 $\mu$m between 2004 and 2006, although this is only based on two data points with high uncertainties. The similar eclipse depths at all wavelengths indicates that even at 4.5 $\mu$m, the ring remains optically thick near central eclipse.

\begin{figure}
\epsscale{1.0}
\plotone{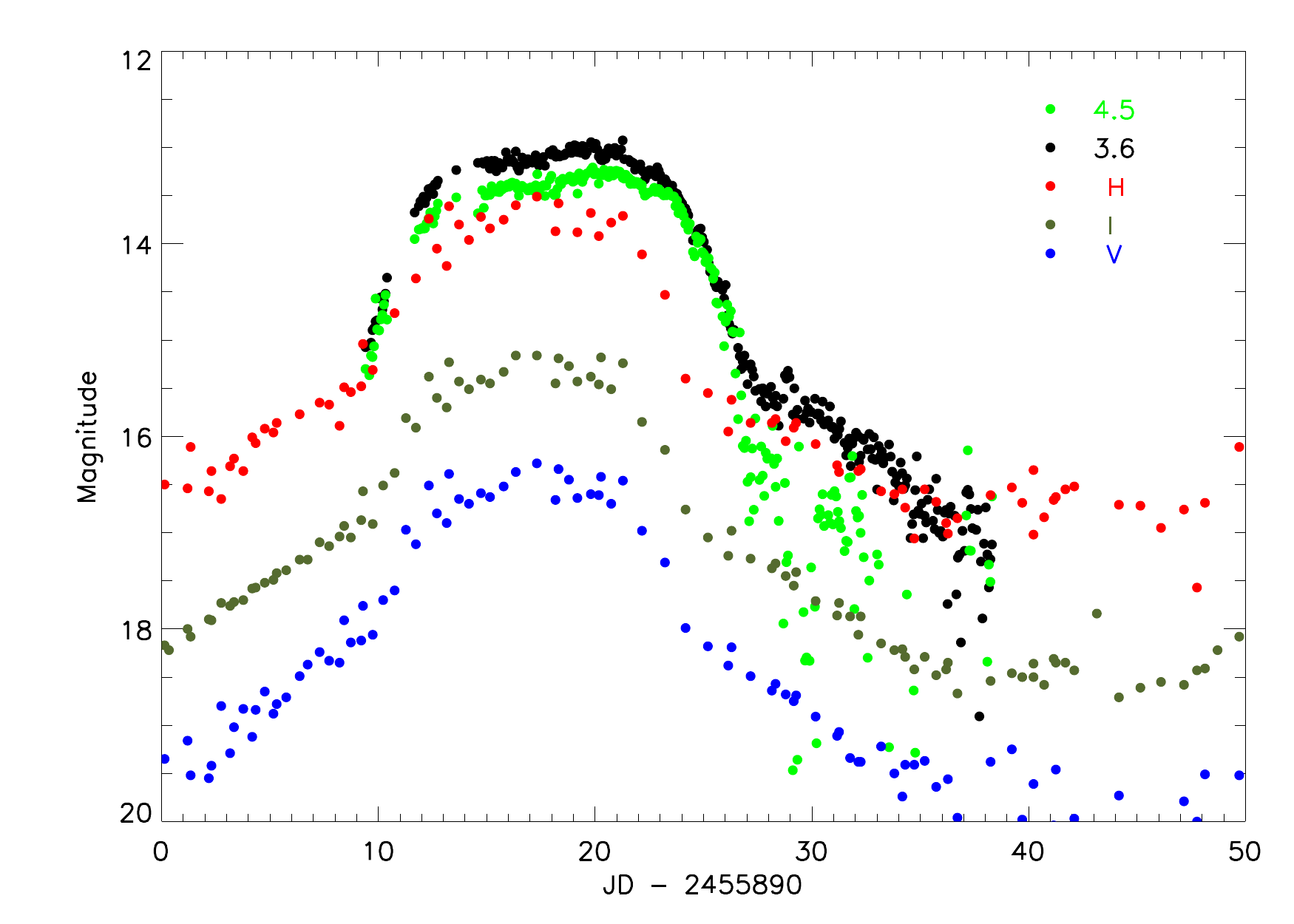}
\caption{The light curve of KH 15D during the CSI 2264 campaign in December 2011. The Spitzer data were all obtained during a single cycle, while the $VIH$ data include one cycle immediately prior and one cycle immediately following. The bright phase extends for a longer period of time in the Spitzer data because we are partially seeing through the obscuring ring edge, especially during egress. The asymmetry in this transparency between ingress and egress is also apparent. Error bars have been omitted for clarity, although the amount of scatter at 4.5 $\mu$m when the system is faint is due to the high uncertainty in those points.}
\label{YSOVAR}
\end{figure}

During the CSI 2264 campaign, high cadence Spitzer photometry was obtained during part of one egress, maximum brightness, and most of one ingress. $VIJH$ band photometry was also obtained at CTIO, often within half a day of the dates on which Spitzer data were collected. By phasing data from the cycle before and the cycle after the Spitzer campaign, it was possible to fill out the $VIJH$ light curves very well. Figure \ref{YSOVAR} shows the resulting light curves for that orbital cycle. It is quite evident that the shape of the light curve at Spitzer wavelengths is rather different from the shape at ground-based wavelengths. In particular, the system brightens a little earlier and more rapidly, and then remains brighter for a substantially longer period of time, beginning its rapid descent (ingress) about two days later than was the case at $VIJH$ wavelengths. We interpret this to mean that the ring edge is somewhat transparent at Spitzer wavelengths, allowing us to see a bit more photospheric emission for a bit longer than at shorter wavelengths. There is more transparency during ingress, indicating that the effect is not identical between ingress and egress.  

\begin{figure}
\epsscale{1.0}
\plotone{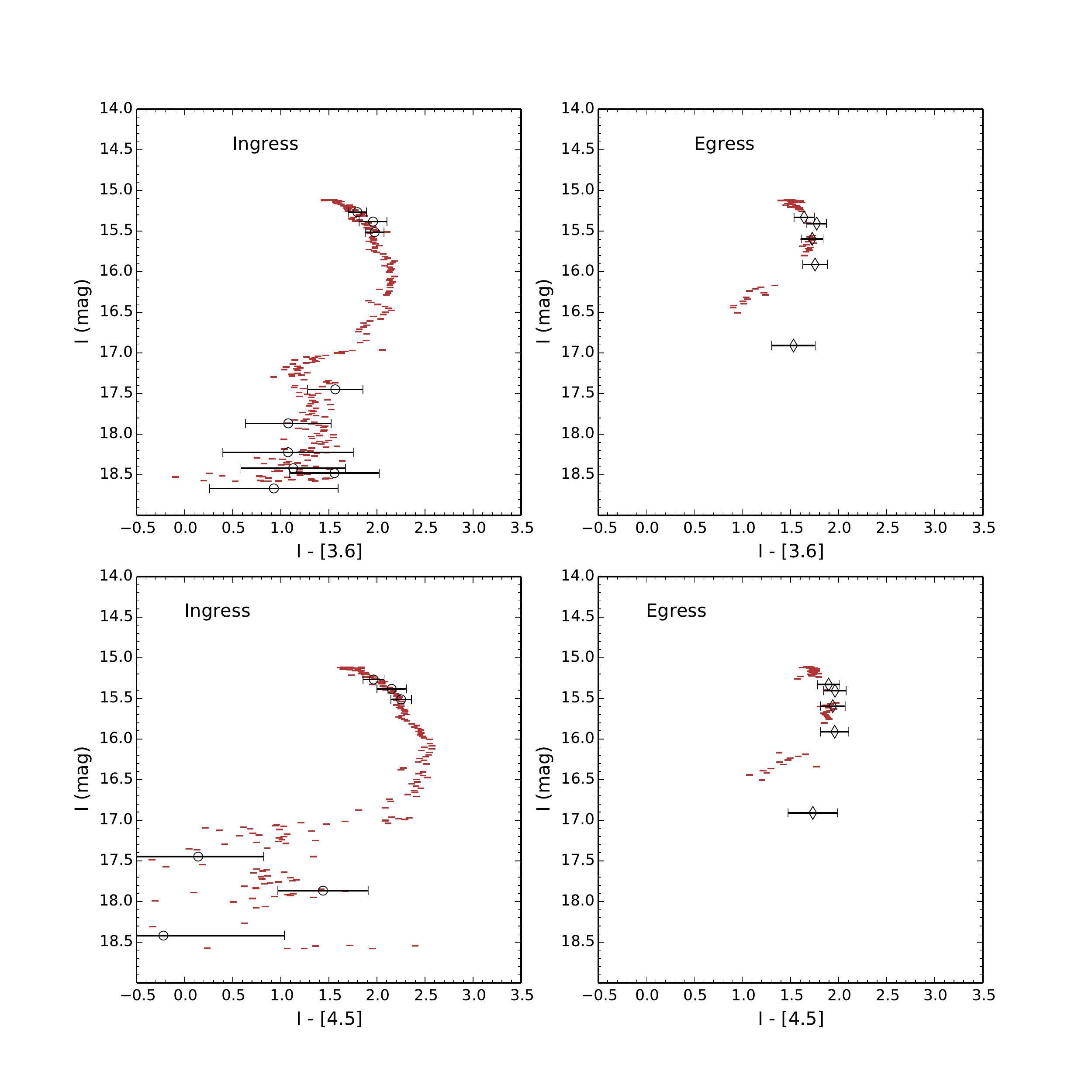}
\caption{The color variations during the CSI 2264 campaign differ between ingress and egress. Here we show $I-[3.6]$ and $I-[4.5]$ CMDs for ingress (left) and egress (right). Black diamonds represent nearly simultaneous observations in the optical and near-infrared, while red points were derived by using the actual infrared data and an interpolated value of the \emph{I} magnitude based on the average light curve during that phase as shown in Figure \ref{YSOVAR}. The ring edge appears to be partially transparent at Spitzer wavelengths, since the light reddens dramatically during both ingress and egress. The transparency is also greater at 4.5 $\mu$m, as expected from Mie scattering by grains of a size comparable to the wavelength of the light. Light path differences between ingress and egress, as star B moves along its orbit, presumably account for the different levels of transparency between ingress and egress.}
\label{modelcmd}
\end{figure}

These results can be further illustrated by color-magnitude diagrams based on the $I-[3.6]$ and $I-[4.5]$ colors calculated through interpolation between phase measurements on the I-band light curves. Figure \ref{modelcmd} shows how the $I-[3.6]$ and $I-[4.5]$ colors behave as KH 15D becomes fainter. The system becomes distinctly redder between peak brightness and $I \sim 16$ mag, when star B is fully occulted, before returning to a bluer color. This is much different from the \emph{VIJH} colors, which get bluer due to the scattered light contribution until the phase when star B is fully occulted. 

The reddening implies that some 3.6 and 4.5 $\mu$m starlight is able to get through the edge of the ring material before the screen becomes fully opaque at some height below the edge of the ring. In other words, the edge of the ring is partially transparent to Spitzer wavelengths. We note that it was also apparently partially transparent at K in the last observing season.  The difference between ingress and egress in the CSI 2264 data is quite apparent. The line of sight through the ring has a different geometry during ingress and egress, owing to the different locations of star B in its orbit. This difference presumably accounts for the fact that the CSI 2264 ingress appears to be a little more transparent than the CSI 2264 egress. Similar trends, indicating partial transparency of the ring edge of variable amounts depending on the phase and cycle, are suggested in the data from 2004-2006, but the relatively sparse sampling of the light curve at those epochs makes it difficult to analyze. The color behavior of the system during the 2011 observing run, from which we have very high cadence data, will be discussed in more detail in Section 4. 

\subsection{An SED for KH 15D in its bright phase with Star A fully visible}

The Spitzer data obtained during the 2004-2006 epoch can be combined with ground-based measurements to construct a spectral energy distribution (SED) of the system when star A is fully visible and star B fully occulted. This is useful for comparing KH 15D to other CTTS and WTTS, which we never get to observe with a reduced contribution from starlight. Disk emission is sufficiently strong in CTTS that it appears in the SED of the system at near-IR wavelengths. Table \ref{SEDTable} shows the magnitudes adopted for the SED, as well as the extinction corrections applied. They are based on the average brightness of the system at maximum light during the period 2004-2006, as summarized in Table \ref{VRIJHKData}, the $JHK$ bright phase measurements by \citet{K05} during that era, and the Spitzer photometry obtained at the bright phase from Table \ref{SpitzerData}. The extinction correction is based on the assumption of foreground reddening with a normal extinction law characterized by $E(B-V) = 0.07$ \citep{S97}. The scattered light from star B should have only a minor effect on the SED at this time, and any light from a putative luminous giant planet would be totally negligible.

\begin{figure}
\plotone{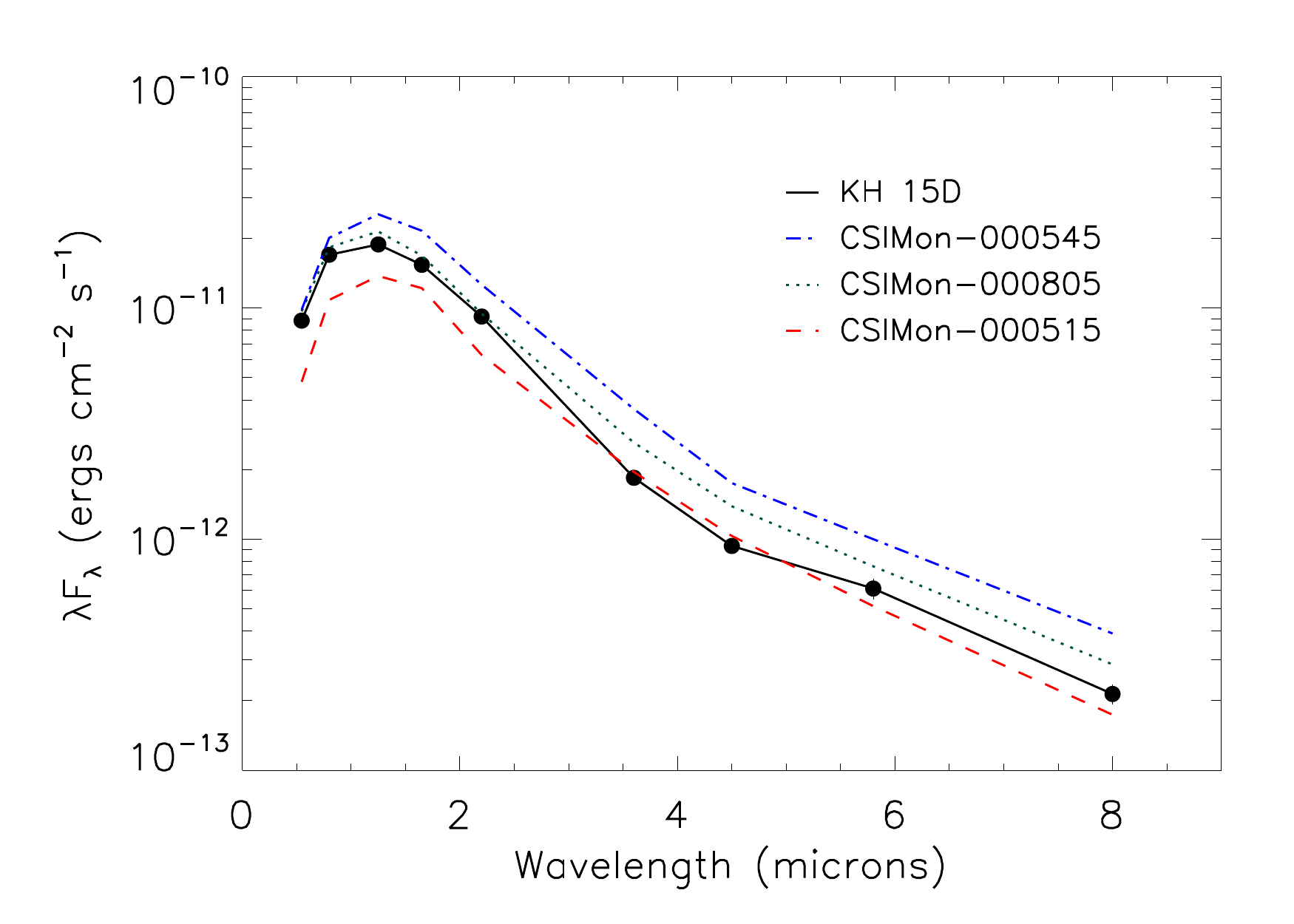}
\caption{The Spectral Energy Distribution of KH 15D during its bright phase prior to 2007, when star A was fully visible  and star B was fully obscured. A comparison with three K7 Class III (WTTS) in NGC 2264 from \citet{C14} is shown. Formal error bars on the photometric data are comparable to the size of the dots, but do not include possible systematic errors due to the difficulty of correcting for the background, especially at longer wavelengths. The agreement is reasonable given the photometric errors. The slightly steeper decline from 1 to 4 $\mu$m and small bump at longer wavelengths are only marginally significant. Based on the overall slope of the SED from 2.2 to 8.0 $\mu$m and the absence of a significant $[3.6] - [8.0]$ color excess, we confirm the classification of the star as a Class III object.}
\label{SED}
\end{figure}

Figure \ref{SED} shows the bright phase SED for KH 15D when star A was fully visible. For comparison, we show SEDs for three K7 WTTSs in NGC 2264 from the data compiled by \citet{C14}. There is moderately good agreement among these. If anything, KH 15D shows less emission in the near-IR, particularly at $[3.6]$ and $[4.5]$. Perhaps this is related to the binary nature of our object, but the discrepancy is relatively small in light of the difficulties of obtaining accurate photometry as described above. The main point, we believe, is that there is no evidence for near-IR disk emission in the bright phase SED of KH 15D. It is consistent with the previous classification of the star as a Class III (WTTS) object based on its bright phase H$\alpha$ equivalent width \citep{H03}. The slope of the SED between 3.6 and 8.0 $\mu$m on a log-log plot is $\alpha = -2.61$, placing the star well below the cutoff of $\alpha = -1.6$ that is commonly employed to identify a Class II (CTTS) object. Its color index of $[3.6] - [8.0] = 0.13$ is also well below the limit of 0.4 adopted by \citet{CB06} and \citet{CH10} to identify sources with disk emission. 

While the conventional interpretation of a source with this kind of SED is that it is ``diskless," KH 15D is an excellent reminder that this is a poor choice of words. We know that KH 15D has a substantial amount of small particulate matter within a few AUs of the stars and that it continues to accrete, presumably from a gaseous disk, and drive an outflow. Viewed from some other angle or at some other time, KH 15D might be considered incapable of such activity. Given its distance and challenging environment, it has not yet proven possible to secure a definitive detection of the source at sub-mm wavelengths. Hopefully this situation can be clarified in the near future with ALMA observations. \citet{CM04} predict that the source will have excess emission in the 10-100  $\mu$m range coming from the occulting inner ring/disk, but observational attempts to confirm this have so far been unsuccessful because of confusion in the region around KH 15D.

\begin{deluxetable}{cccc}
\tabletypesize{\scriptsize}
\tablewidth{0 pt}
\tablecaption{The Spectral Energy Distribution of KH 15D During its Bright Phase in the Epoch 2004-2006 \label{SEDTable}}
\tablehead{
\colhead{Band} & \colhead{ Observed (mag)} & \colhead{Extinction Corrected (mag)} & \colhead{F$_\lambda$ (erg/s/cm$^2$/micron)}
}
\startdata

V & $16.11$ & 15.89 & $1.62 \times 10^{-11}$ \\
I & $14.47$ & 14.32 & $2.13 \times 10^{-11}$ \\
J & $13.42$ & 13.32 & $1.51 \times 10^{-11}$ \\
H & $12.74$ & 12.66 & $9.31 \times 10^{-12}$ \\
K & $12.47$ & 12.41 & $4.18 \times 10^{-12}$ \\
3.6 & $12.79$ & 12.76 & $5.11 \times 10^{-13}$ \\
4.5 & $12.80$ & 12.77 & $2.08 \times 10^{-13}$ \\
5.8 & $12.49$ & 12.47 & $1.05 \times 10^{-13}$ \\
8.0 & $12.66$ & 12.64 & $2.67 \times 10^{-14}$ \\

\enddata
\end{deluxetable}

\section{Discussion}

\subsection{Inner Clearing and Ring Precession}

The varying potential close to a binary system limits the distance from its center of mass at which material can survive on stable orbits. \citet{BK15} have recently provided a convenient summary of earlier work by \citet{HW99} that quantifies that limit in terms of the orbital eccentricity and masses of the components of the binary. Adopting an eccentricity of 0.6 from \citet{HM05} (see also Winn et al. 2006), and masses for star A and B of 0.6 M$_\odot$ and 0.7 M$_\odot$, respectively \citep{W06}, we find that the critical radius inside of which stable orbits cannot exist around KH 15D is 3.85 AU. This is significantly larger than the previous estimate, based on the theory of \citet{AL94}. Note that the limit applies to matter acted on only by the gravitational force of the binary. A rigidly precessing ring presumably might extend into this zone if the motions of its constituent particles were controlled by whatever force (e.g. self-gravity of the ring particles) maintains the rigid precession.

Precession of material in an inclined circumbinary orbit is driven by the fact that the time-averaged potential is not equivalent to that from a point mass at the center of mass. \citet{BK15} also provide an approximate expression that relates the precession rate of the ascending node of an inclined circumbinary orbit of given ``average" radius to the masses and orbital period of the binary. Adopting the masses of the KH 15D components given above and the orbital period of 48.37 days, one finds a precession rate of $3.0 \times 10^{-11}$ radians per second. This may be translated into a linear velocity by multiplication with the radius of the orbit. For a ring at the innermost stable orbit (3.85 AU), we find a translational velocity of 17.5 m s$^{-1}$, in close agreement with what is measured based on the time to cover star A (and uncover star B), as reported above. 

This analysis shows that the long-term photometric behavior of the system matches expectation based on the dynamics of circumbinary material if the ring is centered at about 4 AU, since it precesses at a rate appropriate to that distance from the center of mass. It may not be coincidental that this is also approximately where the innermost stable orbit of the system lies. The radial extent of the rigidly precessing structure remains unknown. \citet{CM04} estimate from dynamical arguments that the ring width is about the same as its radius, which means that it would stretch from 2 to 6 AU. However, they emphasize that this is not strongly constrained by the theory. The improved data on KH 15D over the past decade strongly supports the basic picture of the system, while modifying somewhat the details. The likely location of the ring moves out a bit from previous discussions and the precession period lengthens to about 6500 years. This makes it even more remarkable and fortuitous that we have been able to observe the phenomena associated with this system during the present era. 

\subsection{Forward Scattering Model}

We have made a first attempt at comparing the photometric data gathered with ANDICAM to the ``fuzzy-edge" ring model described in \cite{SA08}. They assumed that the edge of the ring has an optical depth due to dust that varies as a power-law with distance, $\tau = \left(w/y \right)^{\alpha}$, where $w$ is the scale-length of the edge of the ring and $y$ is the sky-projected position perpendicular to the edge of the ring. We hold $w$ and $\alpha$ fixed at the parameters given in \cite{SA08}, which were fit to the $I$-band data along with a dynamical model for the relative positions of the stars and the edge of the ring as a function of time. The model also included the effects of forward-scattering off the edge of the ring and a constant unocculted flux, $f_{in}$, which is able to reproduce the morphology of the eclipses of the $I$-band light curve in great detail.

Here we extend the model to other wavelengths by assuming: 1) grey opacity in the optical and near-infrared (which is appropriate if the grains are much larger than the wavelength); 2) the forward-scattering peak width, $\sigma_D \left(\lambda \right) = \sigma_D \left(0.8 \mu \text{m} \right) \times \left(\lambda / 0.8 \mu \text{m} \right)$, scales with wavelength; we hold $\sigma_D \left(0.8 \mu \text{m} \right)$ fixed at the value found in \citep{SA08}. We use this model to attempt to explain the color behavior as a function of the flux level; we do not attempt to fit the entire light curve, as this would require revisiting the dynamical model of the system, which we leave to future work.

The three parameters we vary for each wavelength are: $f_{out}$, $f_{in}$, and $\kappa_{as}$ (the ratio of forward scattering to large-angle scattering/absorption). We have only attempted to fit the $V$ and $J$ band data as these are well separated in wavelength, but are still short enough wavelength that the grey approximation might still apply. We converted $f_{out}$ and $f_{in}$ to magnitudes $V_{out, in}$ and $J_{out, in}$. We allowed these parameters to vary, we computed $V$ and $J$ as a function of $y$ in the model, and then we computed $V-J$ versus $V$ from this model. We then fit this to the observed $V-J$ versus $V$ and optimized the parameters with a Markov chain to provide the best-fit model to the color-magnitude measurements.  

\begin{figure}
\epsscale{0.75}
\plotone{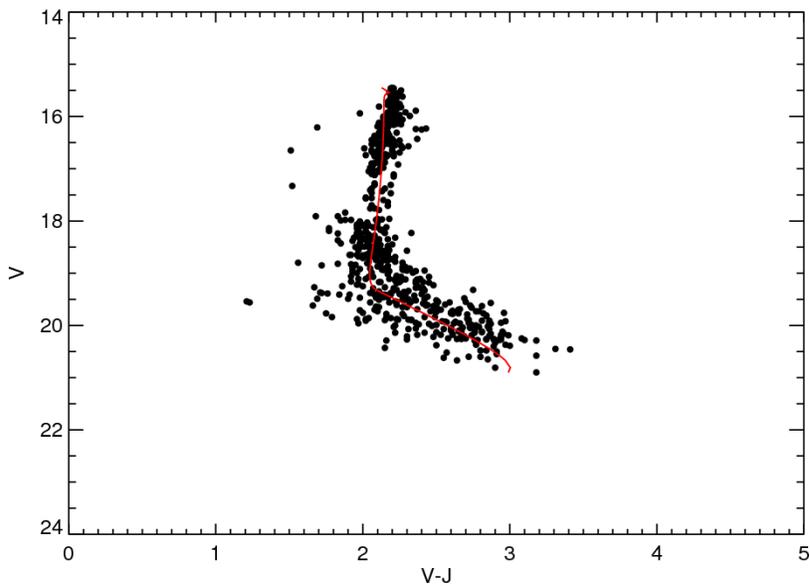}
\caption{The forward scattering model of \cite{SA08} was applied to the $V$ and $J$ band photometry (black points), and the best-fit model (see Table \ref{forward_scat} for optimized parameters) is shown in red. The data agrees with a model of forward scattering by large dust grains, which are responsible for the gradual bluing as star B sets behind the edge of the circumbinary ring. Future work should incorporate a model of the grain size distribution and composition of the dust in order to obtain more accurate opacities and scattering cross sections.}
\label{v_vs_vmj}
\end{figure}

Figure \ref{v_vs_vmj} shows the color-magnitude measurements and the best-fit model. The best-fit parameters are given in Table \ref{forward_scat} (while other parameters were held fixed at the $I$-band values from \citep{SA08}). We also assumed limb-darkening parameters that were independent of wavelength, $\gamma_1 = 0.48$ and $\gamma_2 = 0.21$. 

\begin{figure}
\epsscale{0.75}
\plotone{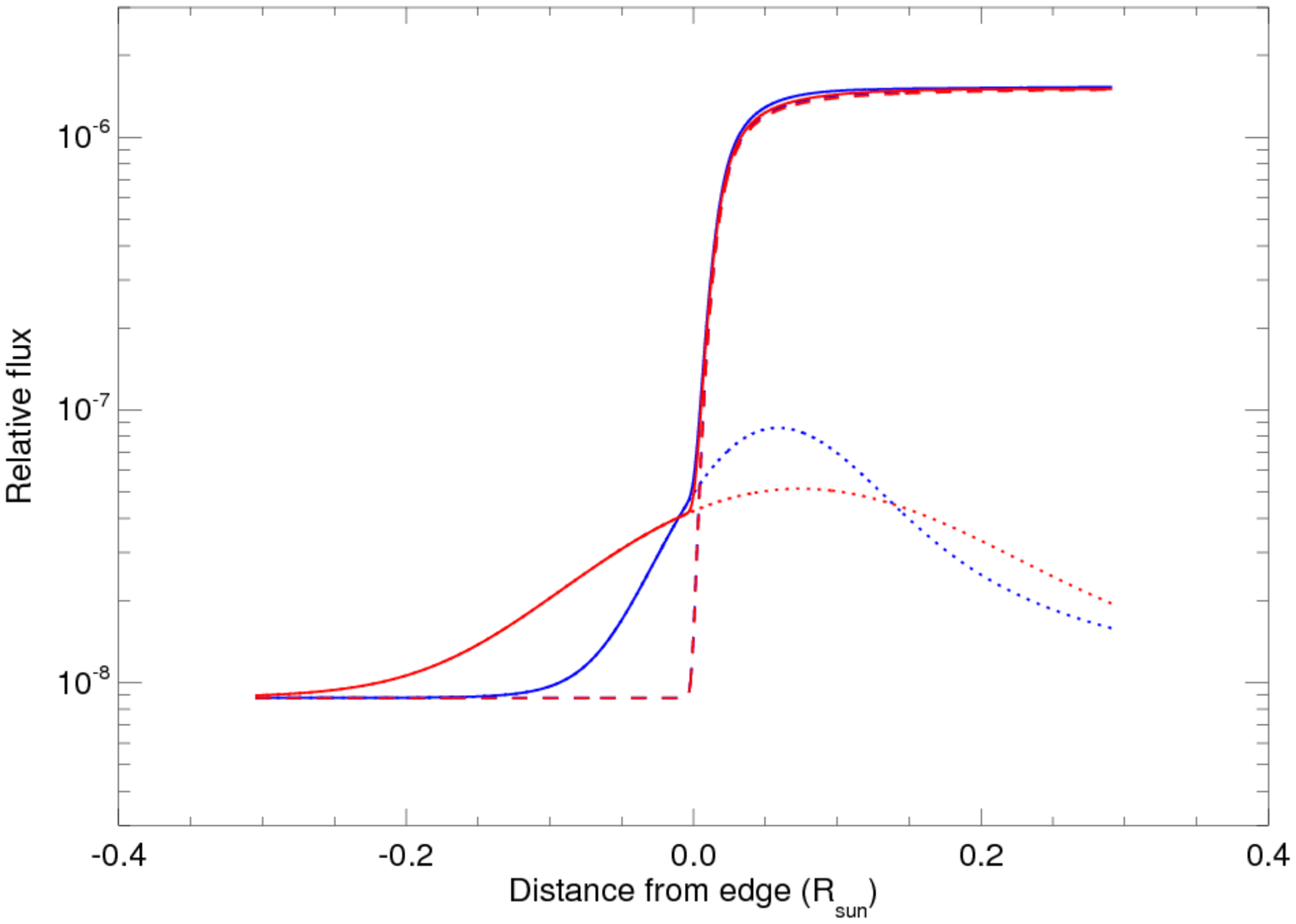}
\caption{The total flux in the $V$ (solid blue) and $J$ (solid red) bands was computed as the sum of the direct (dashed lines) and forward scattered light components (dotted lines). The total flux in the $V$ band remains roughly constant as star B sets (negative distance from edge), while the longer wavelength $J$ band flux decreases, resulting in bluer colors until the distance from the edge is about $-0.1 \, R_{sun}$. As the star sets further below the edge of the ring, the contribution from forward scattered light in the $V$ band declines much more rapidly than the $J$ band component, and the system becomes redder. The $V$ and $J$ band data are in agreement with this model, which attributes the observed colors to forward scattering by large dust grains and decreased optical depth at the edge of the circumbinary ring.}
\label{flux_vs_distance}
\end{figure}

The model does a fair job at reproducing the color variation as a function of wavelength, as shown in Figure \ref{v_vs_vmj}. As the star sets behind the edge, the shorter wavelengths have a narrower and taller forward-scattering peak, causing the sum of the transmitted and forward-scattered flux to first become blue, as shown in Figure \ref{flux_vs_distance}. Then as the star completely sets, the shorter wavelength drops off more rapidly, while the longer wavelength still continues to have significant forward-scattered flux. Since the transmitted flux has dropped significantly, the total light becomes red. We take this agreement as further evidence for a gradual decline in optical depth at the edge of the ring, and for forward-scattering by large dust grains.

Further modeling should include a physical model for the size distribution and composition of the dust, which can give a more accurate model for the opacity and differential scattering cross section. The dynamical model for the system should be revisited, providing constraints on the width of the ring, the shape of the ring, the precession rate, and the variation of the optical depth at the inner and outer edges of the ring. 

\begin{deluxetable}{ll}
\tabletypesize{\scriptsize}
\tablewidth{0 pt}
\tablecaption{Best-Fit Parameters from Forward Scattering Model \label{forward_scat}}
\tablehead{
\colhead{Parameter} & \colhead{Value} 
}
\startdata

$R_\star$ & 1.08 $R_{\odot}$ \\
$V_{in}$ & 21.05 \\
$V_{out}$ & 15.52 \\
$w$ (m) & $1.3 \times 10^9$ \\
$\sigma_{D,V}$ (m) & $2.96 \times 10^9$ \\
$\kappa_{as, V}$ & 2.88 \\
$\alpha$ & 1.5 \\
$J_{in}$ & 21.09 \\
$J_{out}$ & 15.55 \\
$\sigma_{D, J}$ (m) & $6.71 \times 10^9$ \\
$\kappa_{as, J}$ & 2.82 \\

\enddata
\end{deluxetable}

\subsection{Ring Transparency}

Figure \ref{modelcmd} shows that the $I-[3.6]$ and $I-[4.5]$ colors become redder as the system gets fainter during the CSI 2264 campaign. The reddening is present from maximum brightness until $I \sim 17.0$, when the system rapidly transitions to a bluer color. In addition, close inspection of the Spitzer colors reveals that they behave differently at ingress and egress. The following analysis will focus on that set of observations, since it is by far the most comprehensive. While Spitzer photometry was obtained every $\sim$0.1 JD during the run, the \emph{I} band observations were conducted less frequently and actually contain a several-day gap. To fill in the I magnitudes we added phased data from the cycle previous to and the cycle following the actual Spitzer cycle, when computing the I light curve. This adds some uncertainty to the analysis but not enough to obviate the main point revealed in Fig. \ref{modelcmd}. 

Although the inclusion of $I$ band photometry from additional cycles increased the number of phases for which $I-[3.6]$ and $I-[4.5]$ colors could be measured, there were still many Spitzer data points without corresponding $I$ magnitudes. However, there was now enough photometry to determine how the magnitude of the system varied with phase. The data were split into two sections (bright and faint), and a univariate spline interpolation was applied to each. The resulting functions were used to estimate an $I$ magnitude for each phase with Spitzer photometry. The shape of the light curve at $I > 17$ in the second cycle after the Spitzer data were collected was quite different from previous observations. These points were omitted from the interpolation. The estimated magnitudes for phases between $\sim$0.35 and 0.38 are brighter than the measured magnitudes, so the corresponding colors will have a higher uncertainty than those calculated at other phases.

Figure \ref{modelcmd} compares the measured $I-[3.6]$ and $I-[4.5]$ colors (shown in black) to those calculated using interpolated $I$ band magnitudes (shown in red). The estimated colors agree with the observed values, except during egress when the interpolation is more uncertain. Both the photometry and the estimated colors show that the shapes of the CMDs are much different at ingress than they are at egress. As the system becomes fainter at ingress, its colors stay redder than the photosphere of the star until $I \sim 17$, which corresponds to the phase when star B completely disappears behind the screen. The system then becomes bluer until $I \sim 17.5$. At egress, however, the colors only become slightly bluer as the system becomes brighter, and only slight reddening is seen. This implies that the occulting material is somewhat more optically thick at egress than at ingress, which can be explained by the nearly edge-on orientation of the orbital path of the stars. Star B is further away from us at egress, and its starlight must travel a different path through the dust ring to reach us at this phase.

Figure \ref{ANDICAM_cmd} shows that KH 15D maintains fairly constant \emph{V-R} and \emph{V-I} colors before becoming slightly bluer at the phases when star B first disappears behind the edge of the screen. The slope of the bluing increases for \emph{V-J} and \emph{V-H}, but no reddening is observed until minimum light in any of the four colors. This implies that starlight is not able to pass through the ring material, which is completely opaque at these wavelengths. Forward scattered light, which causes the bluing, is the only contribution to the system's brightness when the stars are not directly visible. However, the system behaves differently in \emph{V-K} and at the slightly longer Spitzer wavelengths (3.6 and 4.5 $\mu$m), becoming redder as it becomes fainter instead of remaining at its original color (see Figures \ref{VK_cmd_full} and \ref{modelcmd}).

The reddening in the $I$ vs. $I-[3.6]$ and $I$ vs. $I-[4.5]$ CMDs is consistent with the behavior expected for extinction by grains with sizes on the order of the wavelength of the observed light. To roughly quantify the opacity ratios between the Spitzer band passes, we compared the data to a simple model. For extinction by dust grains, we expect that the flux reaching us through the ring material will follow the relation $F_{\lambda} = F_{\lambda,0} \text{e}^{-\tau_{\lambda}}$, where $F_{\lambda,0}$ is the original flux from the star and $\tau_{\lambda}$ is the wavelength-dependent optical depth. It can be assumed that $\tau_{\lambda} \sim \infty$ across the \emph{VRIJH} bands, appropriate to completely opaque occulting material.

When star B is partially occulted, the flux we detect has a component that is directly visible in addition to the reddened component. There is also some amount of forward scattered light present in the system at all phases, which will be set constant as a first approach. The flux observed on a given night, $F_t$, is therefore the sum of the scattered $\left(F_s \right)$ and non-scattered $\left(F_{\ast} \right)$ components. If we assume $f_d$ is the fraction of the stellar surface that is directly visible at a given phase, $F_t$ may be written as

\begin{equation}
F_{t} = F_s + f_d F_{\ast} + \left(1 - f_d \right) F_{\ast} \text{e}^{-\tau_{\lambda}}.
\label{totalflux}
\end{equation}

The \emph{I} magnitudes may be used to estimate $f_d = \frac{F_t - F_s}{F_{\ast}}$, since starlight in this band apparently does not penetrate the optically thick ring material. For this calculation, we need to know how bright the system would be if the entire surface of star B were directly visible. For the \emph{I} band, we used the value $I_{\ast} = 14.14 \pm 0.02$ that was measured during the 2014-2015 observing season. The $[3.6]_{\ast}$ and $[4.5]_{\ast}$ magnitudes were estimated from the bluest colors in the CMDs, under the assumption that direct light from star B dominates the contribution from reddened light until a substantial fraction of its surface is occulted by the ring material. During egress, the reddening is reduced and the system reaches bluest colors of $I - [3.6] \sim 1.8$ and $I - [4.5] \sim 1.6$ as it becomes brighter (see Figure \ref{modelcmd}). We expect this trend to hold until star B has emerged completely, implying Spitzer magnitudes of $[3.6]_{\ast} \sim 12.34$ and $[4.5]_{\ast} \sim 12.54$ for the fully visible star.

Values for $m_s$ were obtained by the same method used to extract $m_{\ast}$ from the CMDs. We can estimate $I_s - [3.6]_s \sim 1.9$ and $I_s - [4.5]_s \sim 1.6$, assuming $I_s \sim 17.0$ when the entire surface of star B becomes fully occulted. The flux from scattered light in the two Spitzer bands can then be estimated from $[3.6]_s \sim 15.1$ and $[4.5]_s \sim 15.4$.

The fraction of the stellar surface that is directly visible is not necessarily equivalent to the fraction of the total flux received. Limb darkening effects can result in less direct light reaching us when a fraction of the limb is revealed than when the same fraction emerges from a location close to the center. Assuming the surface of the star is circular, we can obtain the limb-darkened flux ratio as 

\begin{equation}
f_{ld} = \begin{cases}
{\displaystyle{C \times \left[2 \int^{y_{max}}_0 \int^{\sqrt{1 - y^2}}_0 1 - u \left(1 - \sqrt{1 - x^2 - y^2} \right) dx dy 
\atop
\hfill + 2 \int^0_{y_{min}} \int^{\sqrt{1 - y^2}}_0 1 - u \left(1 - \sqrt{1 - x^2 - y^2} \right) dx dy \right]}} & : f_d > 0.5 \\ 
C \times \left[2 \int^1_{y_{min}} \int^{\sqrt{1 - y^2}}_0 1 - u \left(1 - \sqrt{1 - x^2 - y^2} \right) dx dy \right] & : f_d < 0.5 
\end{cases} \label{fld}
\end{equation}
where $-1 \leq x, y \leq +1$ and $u$ is the wavelength-dependent limb-darkening coefficient. The values $u_{[3.6]} = 0.2069$ and $u_{[4.5]}$ were adopted from \cite{C13}. In this coordinate system, we have defined $\left(x, y \right) = \left(0, 0 \right)$. If $f_d$ is greater than 0.5, the entire top half of the stellar surface is directly visible. The limits of integration are then $y_{max} = 1$ and $y_{min} = -2 \times \left(f_d - 0.5 \right)$. If $f_d$ is less than 0.5, $y_{max} = 1$ and $y_{min} = 1 - \left(2 \times f_d \right)$. The constant $C$ is a normalization factor, which depends on $u$ as 

\begin{equation}
\frac{1}{C} = 4 \int^1_0 \int^{\sqrt{1-y^2}}_0 1 - u \left(1 - \sqrt{1 - x^2 - y^2} \right) dx dy = \pi \left(1 - \frac{u}{3} \right)
\label{C}
\end{equation}

If the optical depth of the ring material were constant over the range of a stellar diameter, we would expect to see linear reddening with decreasing \emph{I} mag until the entire surface of star B becomes occulted by the ring at $I \sim 17$. The data shown in Figure \ref{modelcmd} indicate that the optical depth actually varies over shorter distance scales. We can extract a function for the optical depth as a function of star B's position with respect to the edge of the ring material by using the interpolated/measured \emph{I} and Spitzer magnitudes to estimate the flux coming from the system in the form of Equation \ref{totalflux}. We can then use the measured $[3.6]$ and $[4.5]$ magnitudes to solve for the optical depth, $\tau$, as a function of the visible fraction, $f$, of star B.

\begin{figure}
\epsscale{0.75}
\plotone{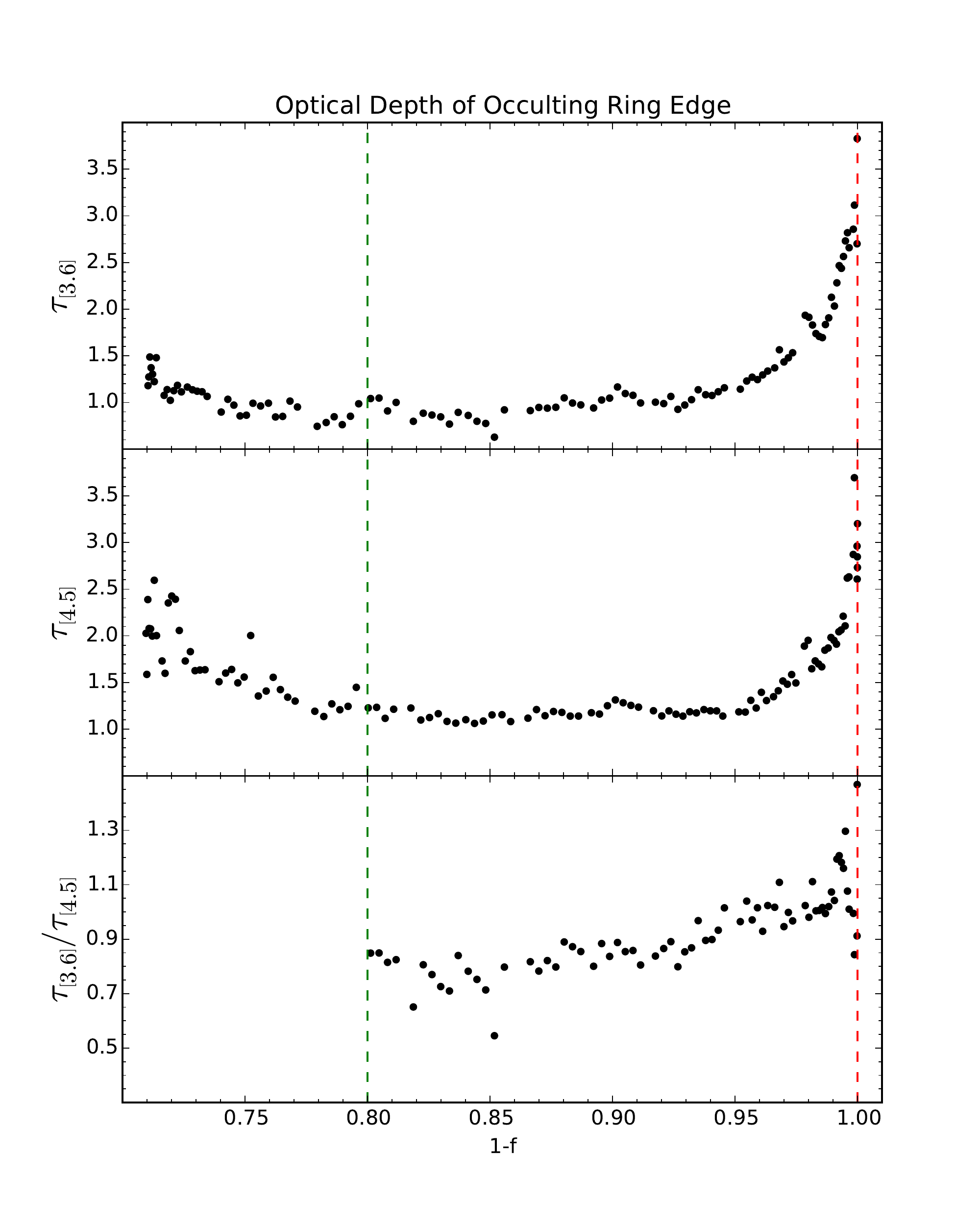}
\caption{Equation \ref{totalflux} can be used along with the magnitude equation to calculate the optical depth of the system at 3.6 and 4.5 $\mu$m as a function of the directly visible fraction, $f$, of the surface of star B. The red, dashed lines indicate where star B is fully occulted around $I \sim 17$ mag. Figure \ref{modelcmd} implies that $\tau_{[3.6]}$ and  $\tau_{[4.5]}$ should increase as the star sets behind the ring material. Our estimates for $F_{\ast}$ are more uncertain than for $F_s$, so the figure is most reliable for $1-f \geq 0.80$ (green, dashed lines), when $\sim$80$\%$ of the stellar surface has disappeared behind the ring.}
\label{modeltau}
\end{figure}

Figure \ref{modeltau} shows that the optical depth, calculated in this manner, decreases until $1-f \sim 0.80$ before increasing again until star B is fully occulted at $1-f = 1$. However, the initial increase in transparency is not consistent with Figure \ref{modelcmd}. The slope of the reddening curve increases as the system becomes fainter, implying that the average optical depth also increases as the star sets further below the edge of the ring. The increase in optical depth can likely be attributed to the values we selected for $F_{\ast}$. Our estimates for the magnitudes of the system at 3.6 and 4.5 $\mu$m when the entire surface of star B is directly visible are uncertain and could be having a significant impact on the values of $\tau_{\lambda}$ we have derived. The estimated values of $F_s$ are much more reliable, so our extinction model still provides a good estimate of optical depth for $1-f \geq 0.80$. The median value from the bottom panel of Figure \ref{modeltau} is $\tau_{[3.6]} / \tau_{[4.5]} = 1.04$, indicating a substantially greyer extinction law than for the interstellar medium (ISM), consistent with the interpretation that we are seeing extinction of starlight by large dust grains in the circumbinary ring at 3.6 and 4.5 $\mu$m.

Figure \ref{VK_cmd_full} shows that during an eclipse this past observing season, the \emph{V-K} color also becomes redder as the system becomes fainter. This trend had not previously been detected at 2.2 $\mu$m, suggesting that we are looking along a different line of sight to the circumbinary ring than in previous epochs. The \emph{K} band photometry was not obtained with as high of a cadence as the 2011 Spitzer photometry, so it would be more difficult to derive a functional form for the optical depth from this data. However, the reddening at 2.2 $\mu$m can be studied further when ANDICAM observations resume in Fall 2015. It may be that we are just barely seeing through part of the ring edge at $K$-band wavelengths, although whether this will continue to be true is unknown. 

\section{Conclusions}

We have presented the results of optical and infrared observations of the eclipsing T Tauri binary KH 15D. $VRIJHK$ photometry was obtained with ANDICAM on the 1.3 m telescope at CTIO from October 2013-April 2014 and again from September 2014-April 2015. The system was also observed with the Spitzer Space Telescope at several epochs between 2004 and 2006, and in December 2011 as part of a campaign by the CSI 2264 team. Follow-up observations at a lower cadence were then conducted between December 2013 and January 2014 to obtain information at eclipse phases that were not observed during the 2011 run. The data have provided the following information about the behavior of KH 15D:

\begin{itemize}
 
 \item The maximum brightness of the system has increased over the last two years, reaching a peak magnitude of $I = 14.14 \pm 0.02$. This is slightly brighter than the expected \emph{I} band magnitude for star B based on archival photometry \citep{WH14}, implying that star B is now fully emerged at apastron, or close to it. Figure \ref{longterm} shows that the leading edge of the occulting screen took $\sim$4 years to cover the orbit of star A, while the trailing edge revealed the path of star B within the same amount of time. This strongly supports models of a rigidly precessing circumbinary ring. The projected ring edges, leading and trailing, move together across the sky at about 15 meters per second, appropriate for a ring radius of about 4 AU. The precession period in this model is about 6500 years. 
 
\item The $VRIJH$ colors indicate that the ring is opaque at those wavelengths, and that the forward scattered light off the ring edge has a fairly achromatic wavelength dependence characterized by $\sim\lambda^{-0.25}$. This is confirmed by the model of \cite{SA08}, which shows that the component of scattered light at shorter wavelengths begins to decrease at shorter distances from the ring edge than the scattered light at longer wavelengths, resulting in bluer colors before eclipse. 

\item The CSI 2264 high cadence data reveals reddening in the $I-[3.6]$ and $I-[4.5]$ colors that we interpret as slight transparency of the ring edge during that cycle. A simple model suggests that the ratio of 3.6 $\mu$m to 4.5 $\mu$m opacity is about 1.04, consistent with extinction by large grains and smaller than the value of 1.25 expected for a $\lambda^{-1}$ law such as applies in the interstellar medium.  Similar transparency may also be visible in the latest season of $K$-band photometry, but additional data are needed to confirm this. It is clear that the opacity of the ring edge can change even between an ingress and egress of the same cycle, presumably due to the different geometry caused by the stellar orbit in relation to the ring. 
 
\end{itemize}

Observations of dust in protoplanetary disks are normally carried out at much longer wavelengths than 2.2, 3.6, and 4.5 $\mu$m and provide information about cooler grains much further out from the star than is the case for KH 15D. The light from the host star in a typical WTTS system overwhelms the contribution from the disk at these wavelengths, making it impossible to obtain detailed information about solids inside 5 AU within these systems, even though we expect such solids to be present.  The unique orientation of KH 15D continues to provide an opportunity to study the grain properties and their distribution within an inner disk and these data can continue to serve as a guide to the solar nebula during the initial stages of planet formation in the solar system and to planet formation in other low mass stars, particularly binaries.

\acknowledgments
This work is based in part on observations made with the Spitzer Space Telescope, which is operated by the Jet Propulsion Laboratory, California Institute of Technology under a contract with NASA. It has also made use of the 1.3 m telescope operated by the SMARTS consortium at Cerro Toloto Interamerican Observatory in Chile. We are grateful for funding from the CT Space Grant Consortium that partially supported this research. We are grateful to Wesleyan undergraduate student Rachel Aronow for her help with this project.

\begin{appendices}

\section{\\ ANDICAM Data Reduction} \label{App:AppendixA}
\setcounter{table}{0}
\renewcommand{\thetable}{A\arabic{table}}

The magnitude of KH 15D in the ANDICAM images was determined using differential photometry. The optical images contain a subset of three of the seven optical reference stars listed by \cite{H05} as calibrators (labeled C, D, and F). However, \cite{WH14} narrowed the list of reference stars down to two after detecting variability in the brightness of star D. The near-infrared photometry was calibrated using a set of five reference stars (labeled 1, 2, 3, 4, and 7), although two of the five (stars 1 and 7) were faint compared to the background in the $K$ band and therefore were not used to determine $K$ magnitudes \citep{WH14}. 

To make sure the selected reference stars were stable throughout the most recent observing seasons, we calculated the difference in magnitude between two stars in a given band for each night of observations. This difference was simply [F]-[C] for the optical reference stars, where [F] and [C] denote the measured magnitudes of reference stars F and C, respectively, on a given night. For the near-infrared reference stars, differences were calculated with respect to star 4 (e.g. [1]-[4]). We then calculated the standard deviation, $\sigma$, of the data to quantify the night-to-night fluctuations in the magnitudes of the comparison stars. The resulting values are listed in Tables \ref{VRIerrors} and \ref{JHKerrors}. We found that the standard deviations were much less than 0.1 mag for most of the reference stars, making the differential photometry done with these calibrators a reliable way to determine the magnitude of KH 15D. However, star 2 showed more scatter than the others over both observing seasons. While the deviations weren't large enough for us to discard star 2 as a calibrator, its stability should be checked in future observing seasons.

\begin{deluxetable}{cccc}
\tablecaption{Standard Deviation of Reference Stars F and C \label{VRIerrors}
}
\tablewidth{0 pt}
\tablehead{
\colhead{Year} & \colhead{$V$} & \colhead{$R$} & \colhead{$I$} 
}
\startdata
2013/14 & 0.018 & 0.024 & 0.017 \\
2014/15 & 0.020 & 0.015 & 0.025 \\

\enddata
\end{deluxetable}

\begin{deluxetable}{cccccc}
\tablecaption{Standard Deviation in Reference Star $JHK$ Mag Difference \label{JHKerrors}}
\tablewidth{0 pt}
\tablehead{
\colhead{Band} & \colhead{Year} & \colhead{$[1]-[4]$\tablenotemark{a}} & \colhead{$[2]-[4]$} & \colhead{$[3]-[4]$} & \colhead{$[7]-[4]$}
}
\startdata
J & 2013/14 & 0.036 & 0.080 & 0.025 & 0.028  \\
H & 2013/14 & 0.047 & 0.083 & 0.025 & 0.035  \\
K & 2013/14 & - & 0.089 & 0.028 & - \\
J & 2014/15 & 0.030 & 0.086 & 0.021 & 0.025  \\
H & 2014/15 & 0.043 & 0.088 & 0.019 & 0.031 \\
K & 2014/15 & - & 0.087 & 0.029 & -  \\
\enddata
\tablenotetext{a}{$[x]$ represents the magnitude of a near-infrared reference star, where $x$ can refer to star 1, 2, 3, 4, or 7.} 
\end{deluxetable}

\section{\\ Spitzer Data Reduction} \label{App:AppendixB}
\setcounter{table}{0}
\setcounter{figure}{0}
\renewcommand{\thetable}{B\arabic{table}}
\renewcommand{\thefigure}{B\arabic{figure}}

IRAC has a 5.2$'$ x 5.2$'$ (256 x 256 pixel) field of view, corresponding to a native pixel scale of 1.2$\arcsec$ per pixel, so our images of KH 15D include HD 47887, NGC 2264 IRS1, and a portion of the Cone Nebula. The data from all six sets of observations were processed by the Spitzer pipeline and combined into mosaics that are currently available through the archive. The final image products have a resolution of 0.6$'$$'$ x 0.6$'$$'$ per pixel. Photometry on the data from 2004-2006 was originally done by Scott Allen, a visiting REU student from Vassar College, who found that the background near KH 15D showed too much variation to obtain reliable aperture photometry. Instead of calculating an average background value, he fit a polynomial to the background to determine how it varied as a function of pixel. While the internal consistency of the results was much improved, overall, the data set was too sparse at that time to facilitate interpretation of the results.  

The high-cadence data from the 2011 observational run were processed by Rob Gutermuth of the CSI 2264 team. Gutermuth used his own method to create mosaics from the original images, instead of using those produced by the Spitzer pipeline \citep{G09, C14}. The program PhotVis was used to determine the magnitude of KH 15D through aperture photometry. An aperture radius of 2 pixels was chosen, with a sky annulus extending from the edge of the aperture to 6 pixels from its center \citep{G08}. These photometric results are included in the catalogue of photometry produced by the CSI 2264 team, and KH 15D is identified as CSIMon-001370 in that work.  

Photometry for the 2013/14 data was initially done by us directly on the mosaics from the Spitzer pipeline, using the IRAF \emph{phot} commands. We chose an aperture radius of 10 pixels and a sky annulus that extended a distance of 12-20 pixels from KH 15D, as suggested in the IRAC Instrument Handbook. The results we obtained for the 2013/14 images were similar to what was reported by Gutermuth from the 2011 observations. Both resulting light curves had eclipse depths much shallower than observed at shorter wavelengths, but visual inspection of the images indicated that obvious contamination by the inner filamentry jet identified by \cite{T04} and the diffraction spikes from NGC 2264 IRS1 had to be corrected for before any conclusions could be reached. Both of these contaminating sources are extended along the north-south direction and stronger on the north side of the source (see Figure \ref{BrightFaint}). 

While these bright background sources were only present on one side of the sky annulus we selected in IRAF, some of their light spilled over into part of the KH 15D aperture as well. The \emph{phot} command removes the background from the source aperture by subtracting the median value of counts per pixel within the sky annulus, which in this case led to an underestimate of the background at the northern edge of the aperture. \emph{Phot} rejects outliers when calculating the median sky brightness and is typically very reliable. However, the contamination was so severe and asymmetric in this case that a better method of sky subtraction was required to obtain more accurate photometry. 

Instead of calculating a median value of the background within some annulus and subtracting the same level of noise from every pixel, we used an approach similar to Allen's method and determined how the sky brightness varied across the aperture. For each mosaic, we first extracted two copies of a 40 x 40 pixel box with KH 15D at its center. A 10 x 10 pixel square, including the star itself, was removed entirely from the center of the second box. We performed a linear interpolation across each row of the ``empty" square to estimate the amount of background flux as a function of $x$ pixel location. The interpolation was done in the north-south direction to match the orientation of the contaminating features in the images. The center of the image was filled in based on the background values determined by the interpolation. The entire 40 x 40 pixel box was subtracted from the first copy, which still contained KH 15D at its center. The system's total flux was calculated by summing over all pixels within a 6 x 6 pixel aperture centered on the source in the background-subtracted image.  

We tested the background interpolation for several different dimensions of the ``empty" box and found that the resulting photometry was dependent on the parameters we selected for its size. The strength of the contaminating filament increased the further north we looked from KH 15D, so a box larger than 10 x 10 pixels would have more flux at its edge than would be encompassed in the final aperture. This would result in an overestimate of the background in the central pixel once the box was filled in with the interpolated sky brightness. The lower limit on the size of the box was set by the FWHM of KH 15D. A smaller box would allow pixels containing starlight to be included in the interpolation, again resulting in excess sky subtraction. 

\begin{figure}
\epsscale{1.0}
\plottwo{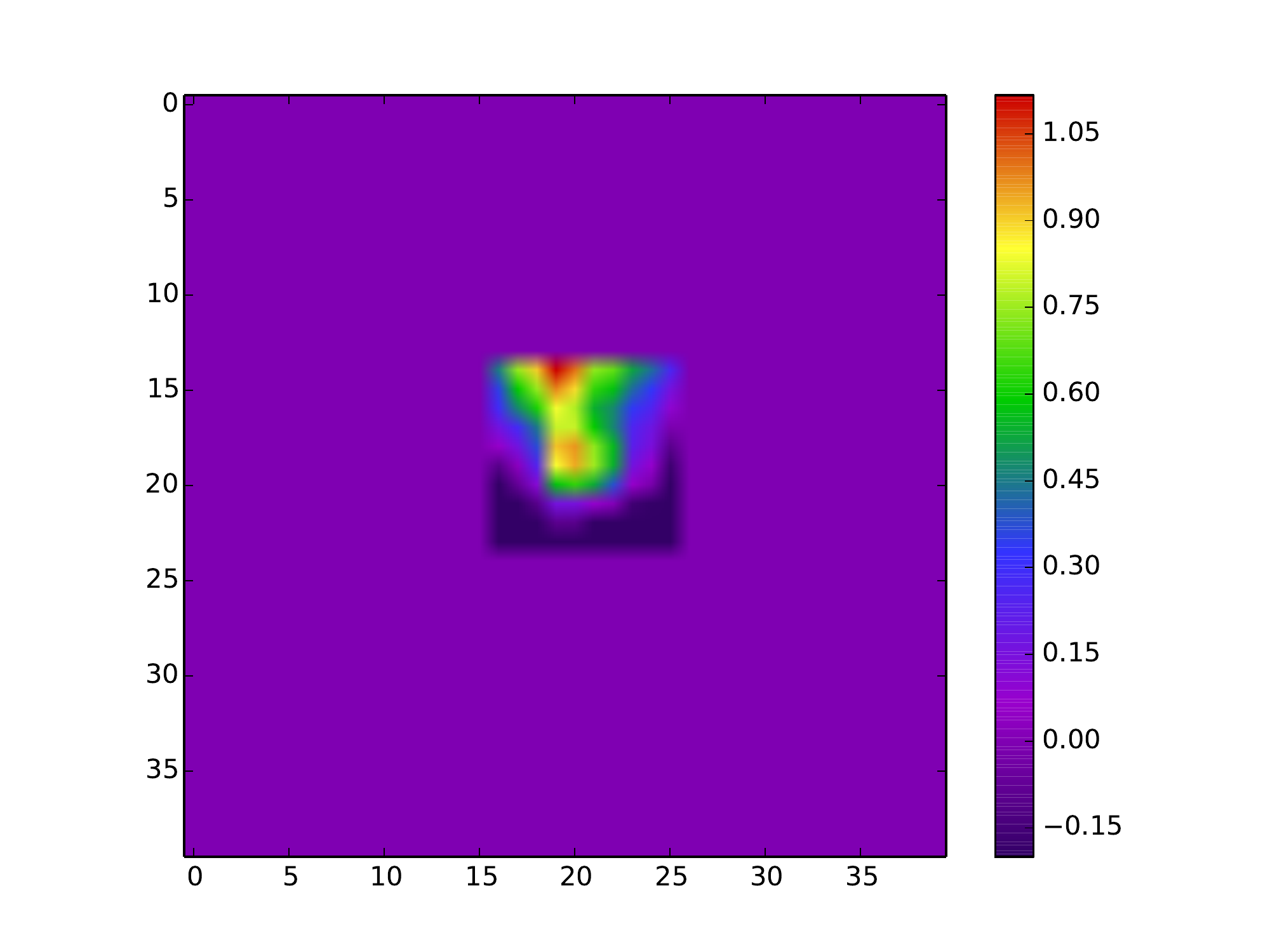}{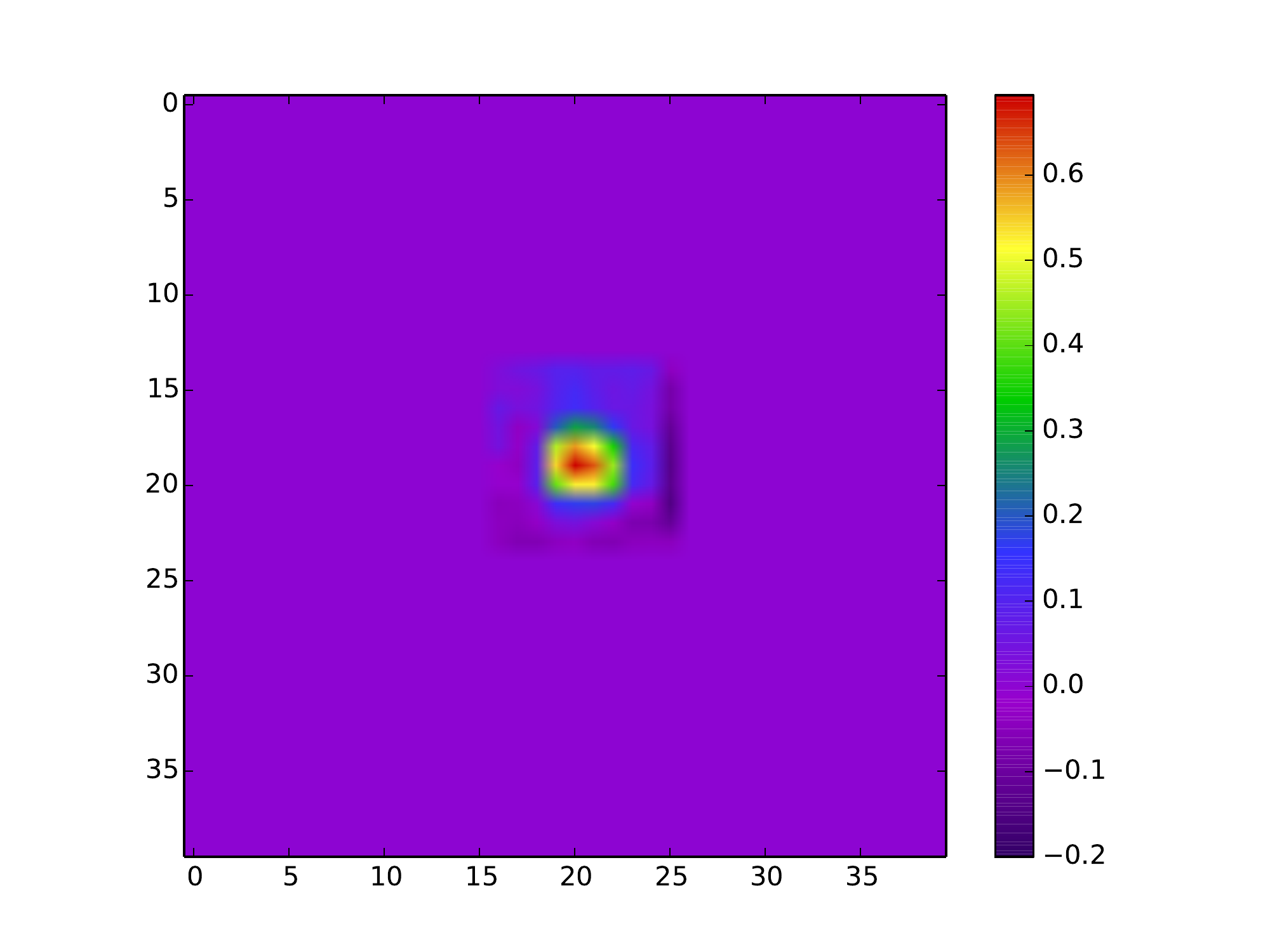}
\caption{The $x$ and $y$ axes above represent pixel locations, while the color bar indicates counts per pixel (relative to the background). The IRAF \emph{phot} command subtracts a median background value from each pixel within a specified annulus (result on left). However, a significant amount of excess flux from KH 15D's jet remains north of the system (North is to the top of the figure; East is to the left). A different method of sky subtraction was required to remove its contribution. We found that the excess flux could be removed by fitting the background with an interpolating function (result on right). The inner box is 6$\arcsec$ on a side. This is an image in the 3.6 $\mu$m band.}
\label{skycomp}
\end{figure}

\begin{figure}
\epsscale{0.75}
\plotone{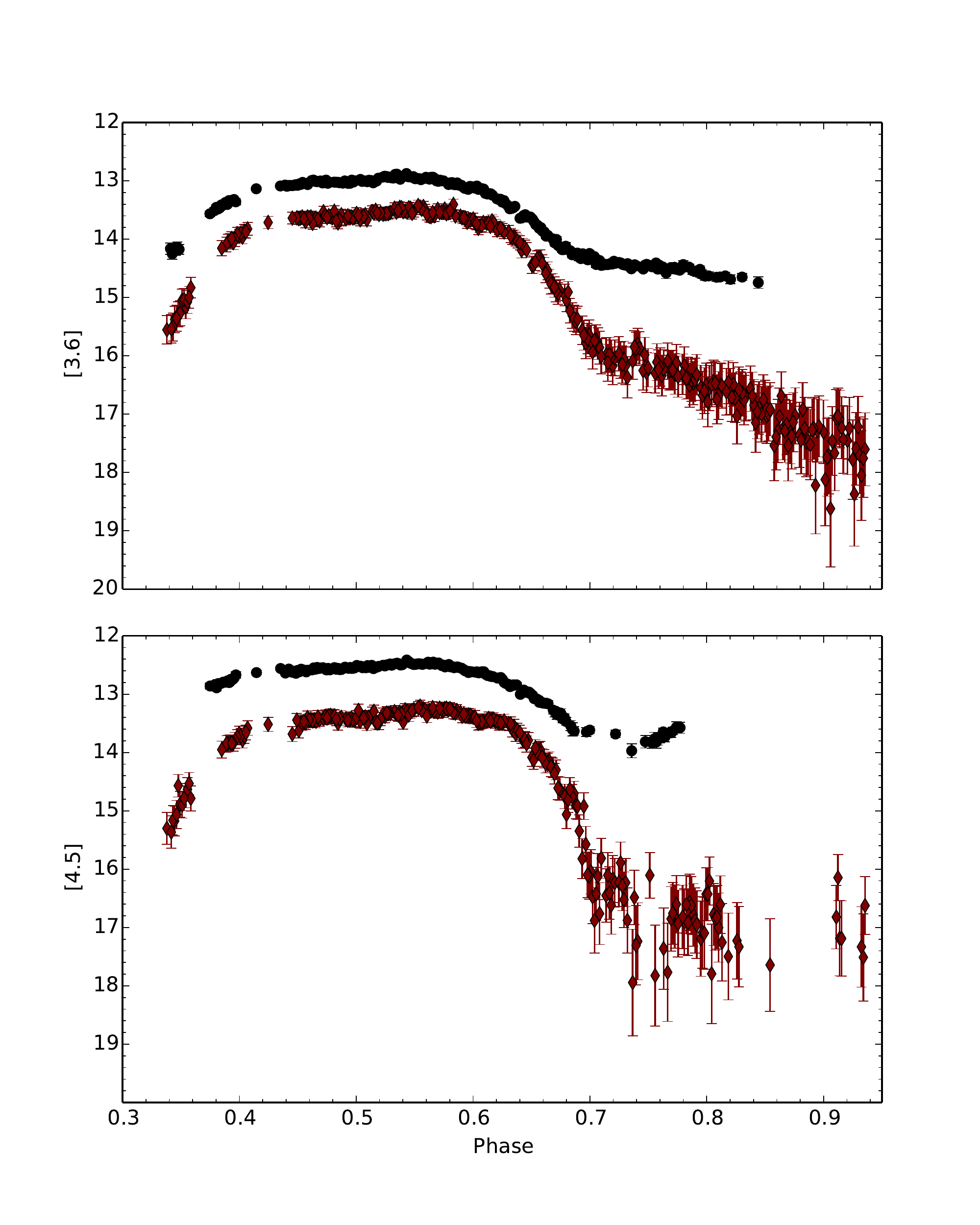}
\caption{The 2011/12 IRAC light curve produced from the photometry obtained with the IRAF \emph{phot} command (black points) showed a much shallower eclipse depth at 3.6 and 4.5 $\mu$m than had been seen at shorter wavelengths. The method of sky subtraction (red points) described in the text shows that the true eclipse depth was greater and more consistent with the change in brightness one can see by eye in the Spitzer/IRAC images. As described in the text, contamination by emission from the diffraction spikes and from the jet are responsible for the difference.}
\label{lightcurvecomp_new}
\end{figure}

Figure \ref{skycomp} compares our method of sky subtraction to the \emph{phot} method for an image taken during the faint phase, when the contrast between KH 15D and the background was lowest. We can see that a significant amount of flux is leftover on the north side of the aperture after subtracting a median background value from each pixel. The residual sky contribution was better removed when we applied the interpolation method instead. We are therefore confident that our sky subtraction method sets a more accurate level for the background contamination in each image than the IRAF aperture photometry. Figure \ref{lightcurvecomp_new} compares our photometry with that provided in the CSI 2264 data base. One can see the offset that arises because of our correction for the asymmetric background contamination. As expected, the offset grows dramatically as the star fades, resulting in a much greater eclipse depth from our photometry. After correction for background contamination we measure an eclipse depth of $\sim$4 mag, which is consistent with observations at shorter wavelengths. 

We applied our background subtraction method to all of the images obtained with Spitzer at all epochs for consistency. For the 2011 images from the CSI 2264 team, we found that the mosaicking procedure adopted by Gutermuth produced images with slightly lower angular resolution per pixel than those compiled by the Spitzer pipeline. Although this didn't have a noticeable effect during the bright phase, it increased the uncertainty in the faint phase photometry. Hence, we acquired the original Spitzer mosaics and obtained photometry from those instead, using the procedure described above. 

\end{appendices}

\end{document}